\documentclass[twocolumn]{aastex63}

\usepackage{epsfig}
\usepackage{graphicx}
\usepackage{amsthm}
\usepackage{amssymb}
\usepackage{rotating}
\usepackage{multirow}
\usepackage{longtable}
\usepackage{savesym}
\savesymbol{tablenum}
\restoresymbol{SIX}{tablenum}
\usepackage{booktabs}

\defcitealias{2019ApJ...873...51L}{Paper~I}
\defcitealias{2020ApJ...888...98L}{Paper~II}
\defcitealias{2021DeBuizer}{Paper~III}
\defcitealias{2022DeBuizer}{Paper~IV}
\defcitealias{2023ApJ...949...82D}{Paper~V}

\accepted{December 30, 2024 by ApJ}

\shorttitle{Surveying G\ion{H}{2} Regions: VI. NGC~3603}
\shortauthors{De\,Buizer et al.}

\begin{document}

\title{Surveying the Giant \ion{H}{2} Regions of the Milky Way with SOFIA: VI. NGC~3603}

\email{jdebuizer@sofia.usra.edu}

\author[0000-0001-7378-4430]{James M. De Buizer}
\affil{\textit{SOFIA}-USRA, NASA Ames Research Center, Mail Stop 232-12, Moffett Field, CA 94035, USA}

\author[0000-0003-4243-6809]{Wanggi Lim}
\affil{IPAC, Mail Code 100-22, Caltech, 1200 E. California Boulevard, Pasadena, CA 91125, USA}

\author[0000-0003-3682-854X]{Nicole Karnath}
\affil{Space Science Institute, 4765 Walnut Street, Suite B, Boulder, CO 80301, USA}

\author[0000-0003-0740-2259]{James T. Radomski}
\affil{\textit{SOFIA}-USRA, NASA Ames Research Center, Mail Stop 232-12, Moffett Field, CA 94035, USA}

\begin{abstract}
We present our sixth set of results from our mid-infrared imaging survey of Milky Way Giant \ion{H}{2} regions with our detailed analysis of NGC~3603, the most luminous G\ion{H}{2} region in the Galaxy. We used imaging data from the FORCAST instrument on the Stratospheric Observatory For Infrared Astronomy (SOFIA) at 20 and 37\,$\mu$m which mapped the central $\sim$8.5$\arcmin\times 8.5\arcmin$ infrared-emitting area of NGC~3603 at a spatial resolution of $\lesssim$3$\arcsec$. Utilizing these SOFIA data in conjunction with multi-wavelength observations from the near-infrared to radio, including Spitzer-IRAC and Herschel-PACS archival data, we investigate the physical nature of individual infrared sources and sub-components within NGC~3603. For individual compact sources we used the multi-wavelength photometry data to construct spectral energy distributions (SEDs) and fit them with massive young stellar object (MYSO) SED models, and find 14 sources that are likely to be MYSOs. We also detect dust emission from the 3 massive proplyd candidates, as well as from the disk and outflow of the evolved blue supergiant, Sher 25. Utilizing multi-wavelength data, we derived luminosity-to-mass ratio and virial parameters for the star-forming clumps within NGC~3603, estimating their relative ages and finding that NGC~3603 is an older G\ion{H}{2} region overall, compared to our previously studied G\ion{H}{2} regions. We discuss how NGC~3603, which we categorize as a `cavity-type' G\ion{H}{2} region, exhibits a more modest number of MYSOs and molecular clumps when compared to the `distributed-type' G\ion{H}{2} regions that share similar Lyman continuum photon rates.
\end{abstract}


\section{Introduction} 

Massive stars form in giant molecular clouds where they initially tend to be highly embedded and hence visible only in mid-infrared to sub-millimeter wavelengths. Eventually, the central protostar reaches such high temperatures that it begins to produce substantial amounts of Lyman continuum photons. These photons create an ionized region in the star's immediate surroundings, which is bright in centimeter radio continuum emission. In the case of a large cluster of massive stars, or multiple generations of massive star formation, the combined Lyman continuum emission can ionize vast regions within the cluster’s host molecular cloud, resulting in the creation of a giant \ion{H}{2} (G\ion{H}{2}) region. These objects typically have sizes of 3-20\,pc in the infrared, and have Lyman continuum photon rates in excess of $10^{50}$\,photons/s. 

Compared to low mass star formation, less is known about the environments of and processes that govern massive star ($M>8\,M_{\sun}$) formation. However, understanding massive star formation is crucial since massive stars are responsible for the creation and distribution of most heavy elements in a galaxy, and are therefore ultimately responsible for the chemical building blocks necessary for the creation of other stars, planets, and life as we know it. Since G\ion{H}{2} regions are the most intense sites of star formation that exist in regular galaxies like the Milky Way, they offer a unique opportunity to study massive star forming clusters and their environments. 

In a series of papers starting with Lim \& De Buizer (2019; hereafter ``\citetalias{2019ApJ...873...51L}''), we have been studying the infrared properties of Galactic G\ion{H}{2} regions using newly acquired data from the Stratospheric Observatory for Infrared Astronomy (SOFIA), as well as archival data from the Spitzer and Herschel infrared space telescopes. Our source list of 42 bona fide G\ion{H}{2} regions and G\ion{H}{2} region candidates comes from De Buizer et al. (2022; hereafter ``\citetalias{2022DeBuizer}''), which was adapted from the original census of \citet{2004MNRAS.355..899C} who identified G\ion{H}{2} region candidates from published 6 cm all-sky surveys along with infrared data from the Midcourse Space Experiment (MSX) and Infrared Astronomical Satellite (IRAS) archives. The original aim of our survey was to produce 20 and 37\,$\mu$m maps of as many of the G\ion{H}{2} regions from the census as we could with SOFIA, and use that data (along with other multi-wavelength data) to understand their physical properties individually and as a population. However, with the recent cancellation of the SOFIA program, our project will remain incomplete with only 29\% of the total population observed (i.e., 12 of 42 G\ion{H}{2} regions). Nonetheless, we will continue to concentrate on studying the infrared properties of each remaining G\ion{H}{2} region for which we have data, comparing and contrasting each region to those regions previously studied in this series. In this, the sixth paper of the series, we will focus on the well-known G\ion{H}{2} region, NGC~3603.

In terms of Lyman continuum photon rate ($4.1\times10^{51}$\,photons/s; \citetalias{2022DeBuizer}), NGC~3603 is the most powerful G\ion{H}{2} region in the Galaxy. At its heart lies a young, massive OB cluster named HD~97950, which is believed to be one of the few Galactic starburst clusters in the Milky Way, containing more than 70 O stars \citep{1994ApJ...436..183M} as well as 3 Wolf-Rayet stars \citep{1989A&A...213...89M, 1995A&A...300..403H} and possessing a dynamic stellar cluster mass of almost $18000\,M_{\sun}$ \citep{2010ApJ...716L..90R}. This powerful cluster has been carving out a bubble in the molecular cloud hosting NGC~3603, leading to a rather small $A_V$ \citep[$\sim$4.5 mag;][] {1998ApJ...498..278E} along the line of sight to the cluster, as well as creating pillars/elephant trunks in the surrounding molecular cloud. It is believed that this expansion into the surrounding molecular cloud may also be triggering the present star formation occurring in NGC~3603 \citep[e.g.,][]{2011A&A...525A...8R}. Compared to our closest massive star forming cluster, the Trapesium in Orion, the NGC~3603 star cluster has about 100 times more ionizing power. Given its relatively close proximity \citep[7.2$\pm$0.1\,kpc\footnote{For more on this adopted distance see Appendix \ref{sec:appendixdist}.};][]{2019MNRAS.486.1034D}, NGC~3603 is invaluable in that it can be studied in rather fine detail and can be used as a proxy in understanding extragalactic starburst phenomena.

In Section \ref{sec:obs}, we discuss the new SOFIA observations and give information on the data obtained. In Section \ref{sec:overview}, we give more background on NGC~3603 as we compare our new data to previous observations and discuss individual sources and regions in-depth. In Section \ref{sec:data}, we discuss our data analysis, modeling, and derivation of physical parameters of sources and regions. We also discuss our interpretation of these results. Our conclusions are summarized in Section \ref{sec:sum}.

\begin{figure*}[tb!]
\epsscale{1.00}
\plotone{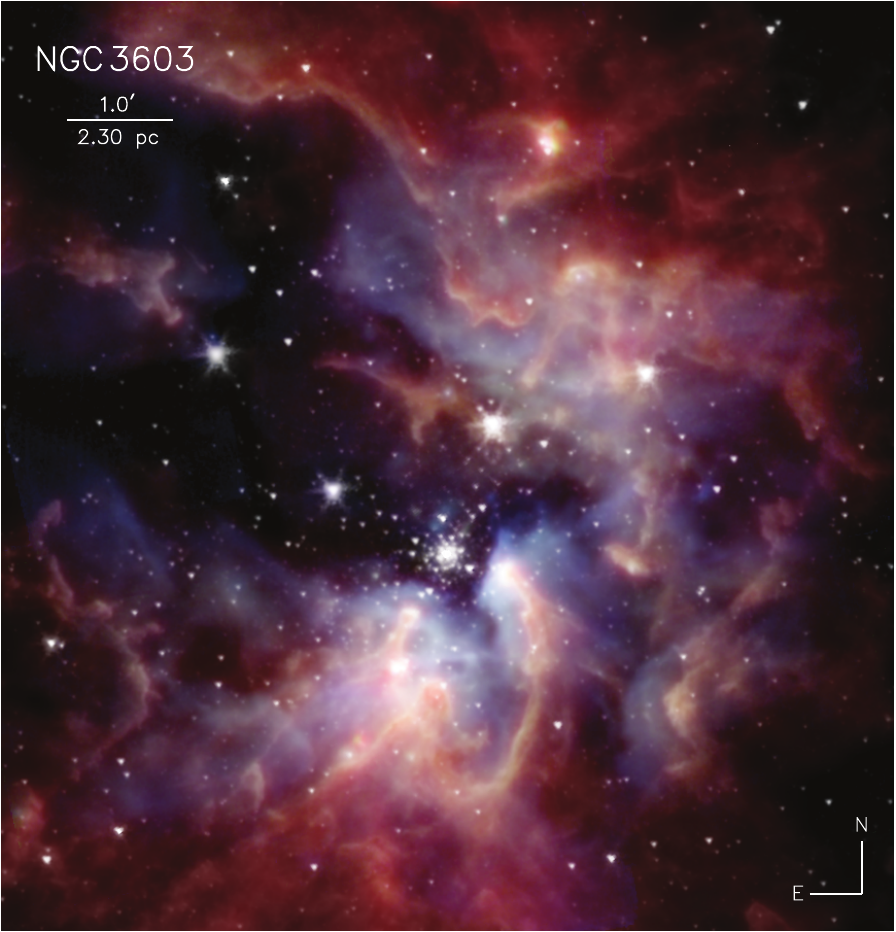}
\caption{ A 4-color image of the central $\sim8\farcm5\times8\farcm5$ (19.6$\times$19.6\,pc) region of NGC~3603. Blue is the SOFIA-FORCAST 20\,$\mu$m image, green is the SOFIA-FORCAST 37\,$\mu$m image, and red is the Herschel-PACS 70\,$\mu$m image. Overlaid in white is the Spitzer-IRAC 3.6\,$\mu$m
image, which traces the revealed stars within NGC~3603, field stars, and hot dust.\label{fig:fig1}}
\end{figure*}

\begin{deluxetable*}{lccrrrrrrl}
\tabletypesize{\scriptsize}
\tablecolumns{9}
\tablewidth{0pt}
\tablecaption{SOFIA Observational Parameters of Compact Sources in NGC3603}
\tablehead{\colhead{  }&
            \colhead{  }&
            \colhead{  }&
           \multicolumn{3}{c}{${\rm 20\mu{m}}$}&
           \multicolumn{3}{c}{${\rm 37\mu{m}}$}&
           \colhead{  }\\
           \cmidrule(lr){4-6} \cmidrule(lr){7-9}\\
           \colhead{ Source }&
           \colhead{ R.A.}&
           \colhead{ Decl. }&           
           \colhead{ $R_{\rm int}$ } &
           \colhead{ $F_{\rm int}$ } &
           \colhead{ $F_{\rm int-bg}$ } &
           \colhead{ $R_{\rm int}$ } &
           \colhead{ $F_{\rm int}$ } &
           \colhead{ $F_{\rm int-bg}$ } &
           \colhead{ Aliases }\\
	   \colhead{  } &
           \colhead{(J2000) }&
           \colhead{(J2000) }& 	   
	   \colhead{ ($\arcsec$) } &
	   \colhead{ (Jy) } &
	  \colhead{ (Jy) } &
	   \colhead{ ($\arcsec$) } &
	   \colhead{ (Jy) } &
	   \colhead{ (Jy) } &
        \colhead{  }\\
}
\startdata
1	&	11:14:46.42	&	-61:15:02.2	&	10  &	43.2	&	20.2	&	9.2	&	41.9	&	30.4 &	\\
2	&	11:14:52.27	&	-61:15:46.8	&	6.9	&	31.5	&	9.21	&	9.2	&	184	    &	73.1 &	\\
3	&	11:14:56.76	&	-61:12:55.5	&	12	&	142	    &	54.9	&	13	&	548	    &	241	 & \\
4	&	11:14:58.36	&	-61:11:26.4	&	4.6	&	5.24	&	2.81	&	4.6	&	40.5	&	23.4 &	\\
5	&	11:14:59.31	&	-61:11:37.2	&	11	&	88.9	&	75.0	&	12	&	408	    &	317	 & \\
6	&	11:15:02.92	&	-61:12:21.7	&	5.4	&	19.2	&	5.01	&	6.1	&	62.2	&	15.6 &	\\
7	&	11:15:03.77	&	-61:15:05.3	&	6.9	&	18.6	&	6.00	&	5.4	&	15.2	&	2.18 &	[NS03] 5A\\
8	&	11:15:07.28	&	-61:14:04.7	&	6.9	&	16.8	&	5.35	&	8.4	&	81.7	&	17.4 &	\\
9	&	11:15:10.16	&	-61:17:37.4	&	6.9	&	49.7	&	7.21	&	7.7	&	188	    &	45.3 &	[NS03] 11\\
10	&	11:15:14.28	&	-61:17:26.2	&	5.4	&	42.7	&	\nodata	&	5.4	&	118	    &	14.8 &	Her-83\\
11	&	11:15:15.28	&	-61:17:36.6	&	7.7	&	65.3	&	20.6	&	9.2	&	256	    &	106	 &  [NS03] 10\\
12	&	11:15:16.54	&	-61:15:01.5	&	6.1	&	15.6	&	10.6	&	6.1	&	8.89	&	2.34 &	\\
IRS4	&	11:15:03.45	&	-61:14:22.3	&	5.4	&	18.2	&	9.33	&	5.4	&	32.2	&	\nodata	& Her-55 \\
IRS9	&	11:15:11.22	&	-61:16:45.2	&	5.4	&	218	    &	110	    &	7.7	&	721	    &	324	& \\
F	&	11:15:05.15	&	-61:16:38.3	&	7.7	&	212	    &	62.7	&	9.2	&	383	    &	91.6 &	\\
P1	&	11:15:12.93	&	-61:15:49.1	&	6.9	&	36.5	&	4.10	&	8.4	&	64.6	&	23.5 &	\\
P2	&	11:15:16.73	&	-61:16:06.3	&	9.2	&	89.3	&	\nodata	&	9.2	&	148	    &	45.9 &	\\
P3	&	11:15:01.29	&	-61:14:46.5	&	9.2	&	38.1	&	\nodata	&	9.2	&	95.5	&	18.3 &	
\enddata
\tablecomments{If there is no $F_{\rm int-bg}$ value for a source, then the source is not well resolved from other nearby sources and/or extended emission. For these sources, the $F_{\rm int}$ value is used as the upper limit in the SED modeling. For sources with alias names, those prefixed with ``[NS03]'' are from \citet{2003AA...400..223N} and those with ``Her-'' are from \citet{2015ApJ...799..100D}.}
\label{tb:SOFIA_compact}
\end{deluxetable*}

\section{Observations and Data Reduction} \label{sec:obs}

The observational techniques and reduction processes employed on the data were, for the most part, identical to those described in \citetalias{2019ApJ...873...51L} for W51A. Below we will highlight some of observation and reduction details specific to these new observations of NGC~3603. For a more in-depth discussion of these details and techniques, refer to \citetalias{2019ApJ...873...51L}. 

Observations were made with the airborne astronomical observatory, \textit{SOFIA} \citep{2012ApJ...749L..17Y}, utilizing the FORCAST instrument \citep{2013PASP..125.1393H}. Data were taken of NGC~3603 across four nights, July $2-5$, 2019 (Flight Numbers $589-592$) with the SOFIA aircraft temporarily deployed to and flying sorties out of Christchurch, New Zealand. All observations were taken at altitudes between 41000 and 43000\,ft, which typically yields precipitable water vapor overburdens of $4-7\,\mu$m at the latitudes where the observations occurred ($-35\arcdeg$ to $-65\arcdeg$). FORCAST is a facility imager and spectrograph that employs a Si:As 256$\times$256 blocked-impurity band (BIB) detector array to cover a wavelength range of 5 to 25\,$\mu$m and a Si:Sb 256$\times$256 BIB array to cover the range from 25 to 40\,$\mu$m. Observations were obtained in the 20\,$\mu$m ($\lambda_{eff}$ = 19.7\,$\mu$m; $\Delta\lambda$ = 5.5\,$\mu$m) and 37\,$\mu$m ($\lambda_{eff}$ = 37.1\,$\mu$m; $\Delta\lambda$ = 3.3\,$\mu$m) filters simultaneously using an internal dichroic. In imaging mode the arrays cover a 3$\farcm$40$\times$3$\farcm$20 instantaneous field-of-view with a pixel scale of 0$\farcs$768 pixel$^{-1}$ after distortion correction. 

All images were obtained by employing the standard chop-nod observing technique used in ground-based thermal infrared observing, with chop throws of up to 7$\arcmin$ and nod throws of up to 16$\arcmin$ in order to be sufficiently large enough to sample clear off-source sky regions uncontaminated by the extended emission of NGC~3603. The mid-infrared emitting area of NGC~3603 is much larger than the FORCAST field of view, and thus had to be mapped using multiple pointings. Each of the eight individual pointings had an average on-source exposure time of about 300s at both 20\,$\mu$m and 37\,$\mu$m. The \textit{SOFIA} Data Pipeline software produced the final mosaicked images (Level 4 data products) from the eight individual pointing images, and these final mosaicked images are presented and used here in this work.

Flux calibration for each source was provided by the SOFIA Data Cycle System (DCS) pipeline and the final total photometric errors in the images were derived using the same process described in \citetalias{2019ApJ...873...51L}. The estimated total photometric errors are 15\% for 20\,$\mu$m and 10\% for 37\,$\mu$m.  

Image quality for SOFIA is typically $\sim$2.5$\arcsec$ (FWHM) resolution at 20\,$\mu$m and $\sim$3.1$\arcsec$ at 37\,$\mu$m. However, on Flights 588-592 there were issues with the chopping secondary mirror, which lead to elongated PSFs in the chop direction, and the amount of elongation varied from negligible to twice the normal FWHM. Chop direction was changed from pointing to pointing, and thus so did the direction of elongation. Therefore, in the final mosaicked image, point sources may appear elongated and in different directions depending on their location within NGC~3603.

The final mosaics also had their astrometry absolutely calibrated using Spitzer data by matching up the centroids of point sources in common between the Spitzer and SOFIA data. Absolute astrometry of the final SOFIA images is assumed to be better than 2$\farcs$0, which is a more conservative estimate than that quoted in our previous papers (i.e., $1\farcs0-1\farcs5$) due to the poorer image quality and because of changes in the focal plane distortion and our ability to accurately correct it with the limited calibration data available for these observations.

\begin{figure*}[tb!]
\epsscale{1.00}
\plotone{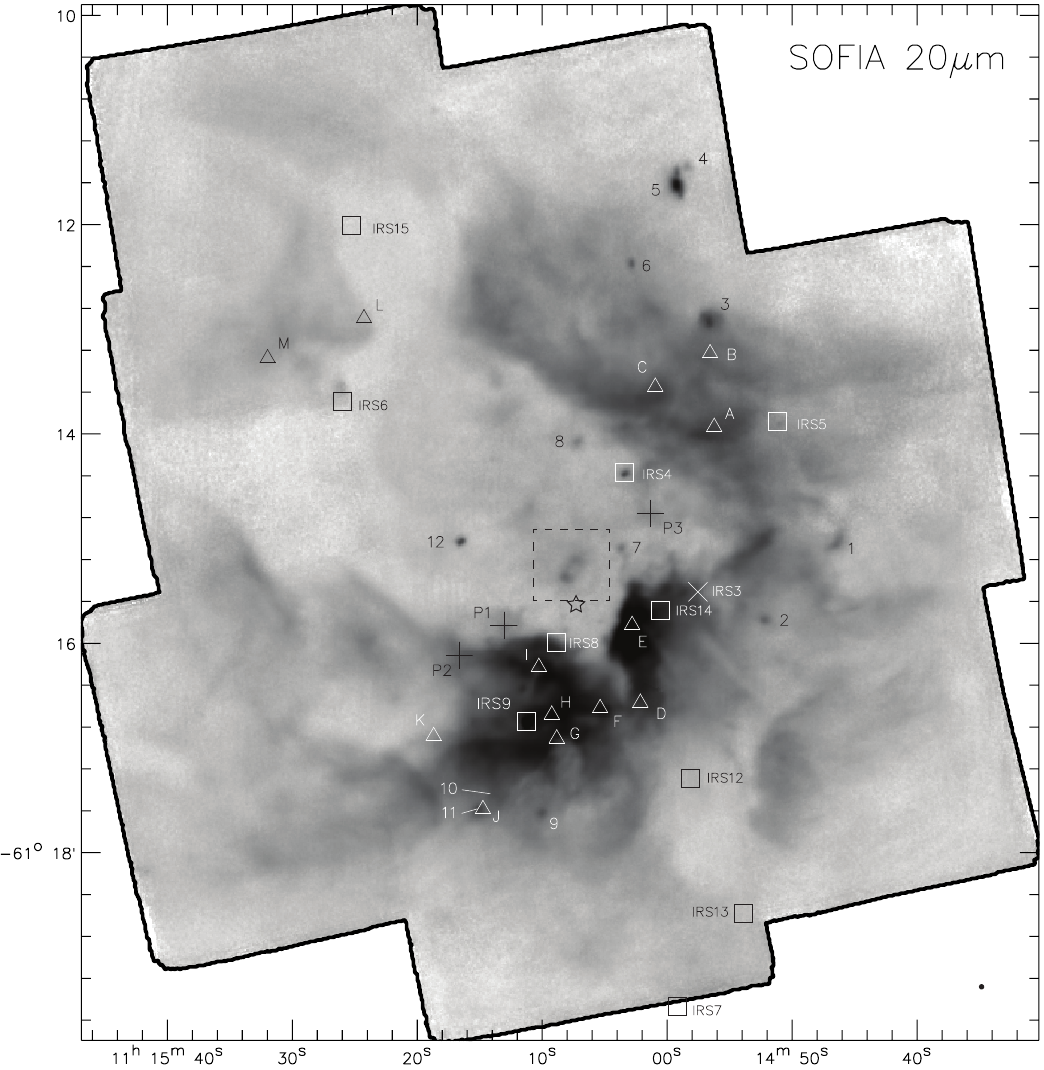}
\caption{NGC~3603 image mosaic taken at 20\,$\mu$m by SOFIA shown in inverse color (i.e. brighter features are darker in color). Sources discussed in the text are labeled: the star show the location of the approximate center of the HD~97590 OB star cluster, the triangles are the locations of radio continuum peaks from \citet{1999AJ....117.2902D}, squares are the locations of previously identified near-infrared sources (see Table~\ref{tb:IRS}), crosses mark the location of the proplyds, and numbers label the mid-infrared compact sources identified in this work. The dashed square surrounds the source Sher 25 and has the approximate field size displayed in Figure~\ref{fig:Sher25}. The black dot in the lower right indicates the resolution of the image at this wavelength.\label{fig:fig2}}
\end{figure*}

\begin{figure*}[tb!]
\epsscale{1.00}
\plotone{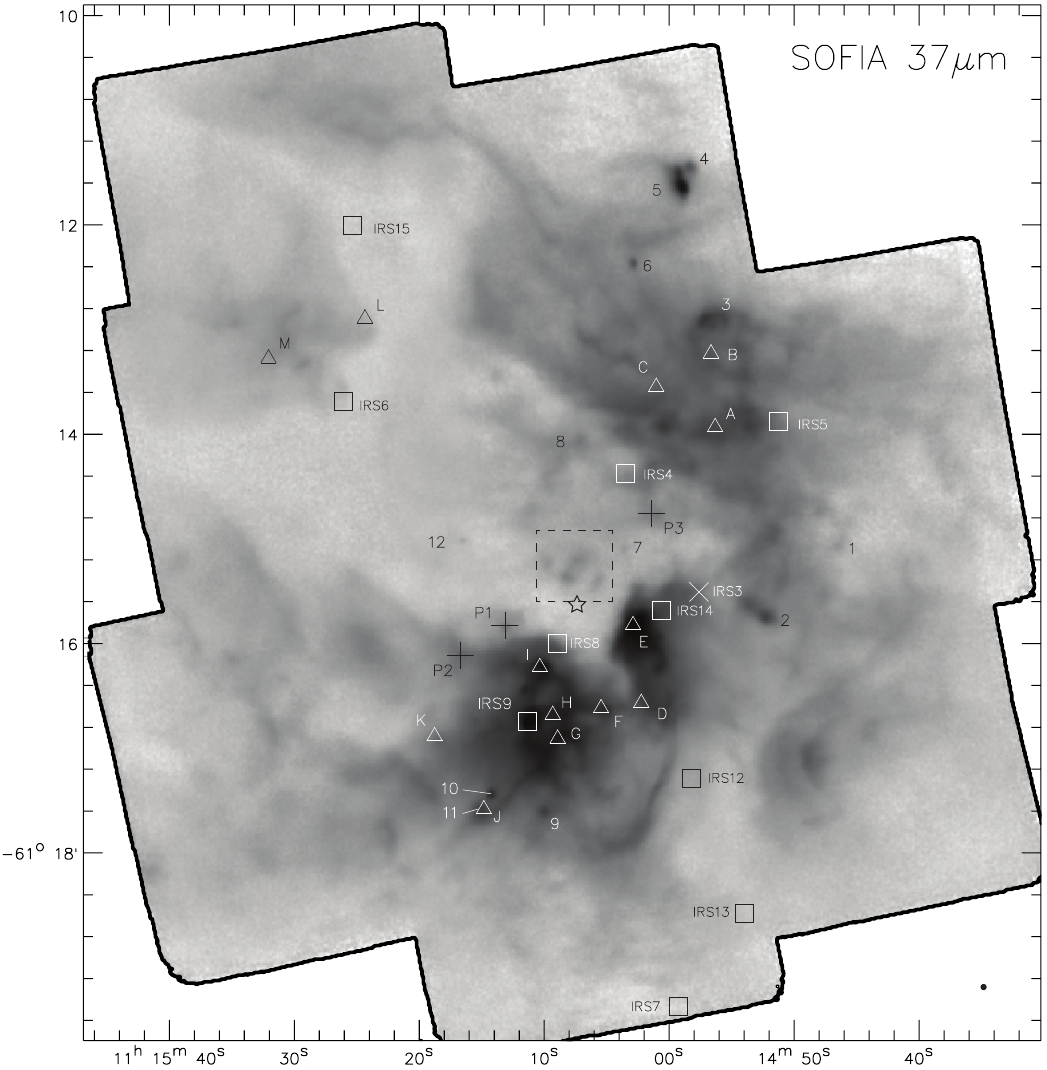}
\caption{NGC~3603 image mosaic taken at 37\,$\mu$m by SOFIA. See caption of Figure~\ref{fig:fig2} for explanation of symbols and figure annotation.\label{fig:fig3}}
\end{figure*}

In addition to the SOFIA data, we also utilize science-ready imaging data from the Spitzer Space Telescope
and Herschel Space Telescope archives. Figure~\ref{fig:fig1} shows a 4-color composite image made from the SOFIA 20 and 37\,$\mu$m data (blue and green, respectively), the Herschel 70\,$\mu$m imaging data (red), and the Spitzer-IRAC 3.6$\mu$m imaging data (white). Figure~\ref{fig:fig2} shows just the 20\,$\mu$m SOFIA final imaging mosaic, and Figure ~\ref{fig:fig3} shows just the 37\,$\mu$m SOFIA final imaging mosaic. 

\section{Comparing SOFIA Images of NGC~3603 to Previous Imaging Observations} \label{sec:overview}

Since most G\ion{H}{2} regions are buried deep within giant molecular clouds, NGC~3603 is somewhat rare in that it is a rather optically bright, which is why observations of this object date back to its discovery by Sir John Herschel in 1834. There is consequently a wealth of optical and near-infrared data on NGC~3603 \citep[e.g.,][]{1969ApL.....4..199G, 1999A&A...352L..69B, 2002A&A...394..253N}, especially concerning the stellar cluster HD~97590 (which can be seen just below the center of Figure~\ref{fig:fig1}) and its immediate environment \citep[e.g.,][]{1989A&A...213...89M, 1998ApJ...498..278E, 2000AJ....119..292B}, including high resolution ($0\farcs2$) Hubble Space Telescope images \citep{1994ApJ...436..183M}. The copious Lyman continuum photons and strong stellar winds from this central stellar cluster can be seen in these optical and near-infrared images to have carved out a large gas-free cavity in the surrounding molecular material. This results in large gaseous pillars on the cavity walls, as well as shaping three compact objects that look like externally irradiated protostellar objects \citep{2000AJ....119..292B,2002ApJ...571..366M} known as ``proplyds'', which were first seen in Orion \citep{2000AJ....119.2919B}. 

The first radio continuum images of NGC~3603 were presented in \citet{1969ApL.....4..199G}, where they showed a single, extended 6\,cm continuum source at $\sim3\arcmin$ resolution peaked on and coincident with the extended H$\alpha$ (656\,nm) nebular emission. Later, the 21\,cm maps of \citet{1980MNRAS.193..261R} at 50$\arcsec$ resolution, resolved the region into two main emission areas, G291.59-0.50 which lies to the north of the revealed massive stellar cluster HD~97950, and G291.63-0.54 which lies to the south. It then took almost until the turn of the twenty-first century for higher angular resolution radio images to appear, with the $\sim$7$\arcsec$ images at 3.4\,cm of \citet{1999AJ....117.2902D} taken with the Australia Telescope Compact Array. With a factor of $\sim$7 times better angular resolution, \citet{1999AJ....117.2902D} identified 13 peaks in the extended radio emission, which were labeled A through M (see labels on Figures~\ref{fig:fig2} and \ref{fig:fig3} which show the positions of the radio peaks). Later observations by  \citet{2002ApJ...571..366M} imaged the central 3$\arcmin$ area at $\sim$1$\arcsec$ resolution at 3 and 6\,cm. 

Similar to the radio continuum observations, there has been a lack of high angular resolution mid-infrared imaging of NGC~3603. \citet{1977ApJ...213..723F} were the first to show 10 and 20\,$\mu$m maps of NGC~3603 which had about 15$\arcsec$ angular resolution, as well as 2.2\,$\mu$m maps with 22$\arcsec$ resolution. They were able to detect the two prominent peaks in the area at 10 and 20\,$\mu$m, the first associated with the E radio peak (which they labeled IRS~1), and the second associated with the F--I radio area (which they labeled IRS~2). A third peak was also seen at 10 and 20\,$\mu$m (named IRS~3), but it is not associated with any radio source. They found several additional 2.2\,$\mu$m point sources (labeled IRS4--15), and using additional J and H data were able to determine that IRS~9, which is the only 2.2\,$\mu$m source found within their 10 and 20\,$\mu$m emission region, was highly reddened. These mid-infrared observations were followed by those of \citet{1982ApJ...255..510L} who mapped only the IRS 1 and 2 regions in the spectral lines of \ion{Ne}{2} and \ion{S}{4} at 7$\arcsec$ resolution.

\begin{figure*}[tb!]
\epsscale{1.17}
\plotone{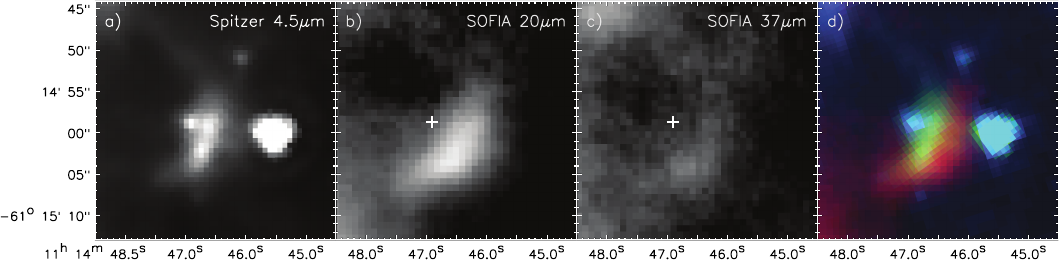}
\caption{Images of source 1 at a) Spitzer 4.5\,$\mu$m, b) SOFIA 20\,$\mu$m, and c) SOFIA 37\,$\mu$m. In panel d) the RGB image contains data from Spitzer 3.6\,$\mu$m (blue), Spitzer 4.5\,$\mu$m (green), and SOFIA 20\,$\mu$m (red). In panels b) and c), the white cross marks the location of the near-infrared stellar source closest to (and presumably heating) the mid-infrared dust arc. \label{fig:Source1}}
\end{figure*}

The next observations in the mid-infrared of NGC~3603 were not until those of \citet{2003AA...400..223N}, who attempted to map the region at $\sim$1$\arcsec$ resolution using the ESO 3.6-m telescope at 12 and 18\,$\mu$m. However, this facility can only chop and nod up to 20$\arcsec$, and the infrared-emitting area of NGC~3603 is much larger. This means their background reference beams were full of emission, and therefore the accuracy of their maps and the photometry of sources were compromised by false structures due to poor background subtraction. Furthermore, unlike the larger coverage of their 12\,$\mu$m map, their shallow 18\,$\mu$m map only covered the IRS1 and IRS2 area. Shortly thereafter in the mid-2000s, Spitzer-IRAC imaging of NGC~3603 generated fine images of the entirety of NGC~3603 in the near-infrared (3.6, 4.5, and 5.8\,$\mu$m), though the mid-infrared 8\,$\mu$m images are plagued with cosmetic issues due to saturation on the brightest point sources on the field. Additionally, the Spitzer-MIPS imaging data at 24\,$\mu$m and longer wavelengths are saturated almost entirely throughout the central 8$\arcmin\times$8$\arcmin$ area \citep{2013MNRAS.430L..20G}. Similarly, images were taken at 3.4, 4.6, 12, and 22\,$\mu$m of NGC~3603 with the Wide-field Infrared Survey Explorer (WISE) satellite, however the two mid-infrared wavelengths (12 and 22\,$\mu$m) were also entirely saturated. The only images obtained of the entire mid-infrared emitting area of NGC~3603 in the last four decades that were unsaturated and unaffected by spurious background subtraction were those from the Midcourse Space Experiment (MSX) satellite, which had a resolution of $\sim$18$\arcsec$ at 21\,$\mu$m \citep[see][]{2010SCPMA..53S.271W}, and the SOFIA 20 and 37\,$\mu$m images at $\sim$3$\arcsec$ resolution presented here.

In terms of large-scale structure, the SOFIA 20 and 37\,$\mu$m images look fairly similar to each other. However, looking at the 3-color image in Figure \ref{fig:fig1} it can be seen that there is a distinct blue haze around and interior to the green and red areas of NGC~3603. Since blue is the 20\,$\mu$m SOFIA image and the shortest wavelength of the three, areas dominated by emission in this filter are likely to be tracing the hotter dust closer to the central stellar cluster. The green and red regions (corresponding to the brighter areas in the SOFIA 37\,$\mu$m and Herschel 70\,$\mu$m data, respectively) trace the colder dust and thus the emission is offset further from the heating of the central cluster. At these wavelengths emission can also be seen coming from the various pillars and trunks, all of which point back to HD~97590. 

\subsection{Discussion of Individual Sources in NGC~3603}

Here we will discuss the individual infrared sources detected by SOFIA within NGC\,3603 and compare them to previous multi-wavelength observations. We will also discuss the nature of the individual sources, where possible. This will often include our findings from the SED modeling that we will describe in detail in Section \ref{sec:cps}, which is based upon the infrared photometry of the SOFIA, as well as Spitzer and Herschel data. Please refer to that section for more detailed information regarding how the SED analyses were performed. 

\begin{figure*}[tb!]
\epsscale{1.17}
\plotone{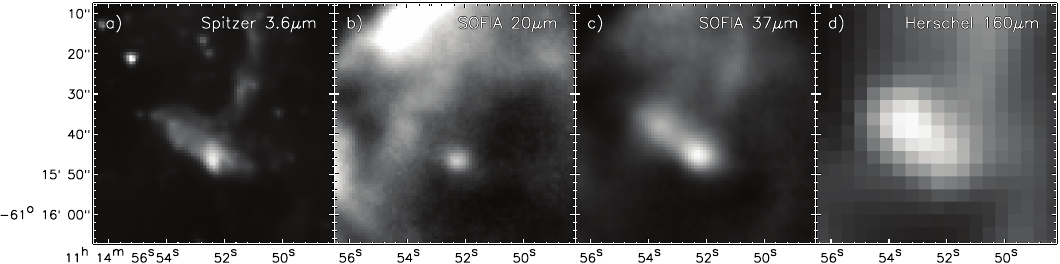}
\caption{Images of source 2 at a) Spitzer 3.6\,$\mu$m, b) SOFIA 20\,$\mu$m, c) SOFIA 37\,$\mu$m, and d) Herschel 160\,$\mu$m. Source 2 is assumed to be the peak seen at 20\,$\mu$m, however there may be a second, more deeply embedded source $\sim$10$\arcsec$ to its northeast (as seen in the 37 and 160\,$\mu$m images). \label{fig:Source2}}
\end{figure*}

\subsubsection{Source 1}
Source 1 appears as a point source in the Spitzer-IRAC 3.6 and 4.5\,$\mu$m images (Figure\,\ref{fig:Source1}) with an arc or ridge of extended ($\sim$6$\arcsec$ long) emission immediately to the west ($\sim$1.5$\arcsec$ away). The point source is presumed to be photospheric emission because has a very faint ($m_G=20$) optical emission component (as seen by GAIA) and is not detected in the IRAC 5.8\,$\mu$m or 8\,$\mu$m images, nor at longer wavelengths. In contrast, the arc of infrared emission becomes the only detected emission source at the SOFIA wavelengths, and is most prominent at 20\,$\mu$m. However, at 37\,$\mu$m the infrared arc begins to fade and in the Herschel 70 and 160\,$\mu$m images (not shown here) there does not appear to be any detectable emission at this location (from either the stellar source or mid-infrared arc) above the widespread nebular emission of the G\ion{H}{2} region. Therefore, the emission seen from $5.8-37.0\,\mu$m is likely from a partial shell or external ridge of emission being heated by the nearby star (seen at $\lambda\le4.5$\,$\mu$m), and is likely not an internally heated YSO. As can be seen in Figure\,\ref{fig:Source1}d, the emission location of the arc moves farther away from the point source location as wavelength increases. This is what is expected for an externally heated object. In fact, the $5.8-37.0\,\mu$m data points are better fit with a simple blackbody of $\sim$205\,K, than with the SED fitting algorithm, which further implies that the mid-infrared source seen with SOFIA is likely not a YSO. This region is not covered by either the \citet{2003AA...400..223N} mid-infrared nor the \citet{1999AJ....117.2902D} radio continuum maps.

\subsubsection{Source 2}
Source 2 appears as an elongated bar of emission measuring $\sim$14$\arcsec$ long running northeast to southwest (p.a.$\sim$300$\arcdeg$). In the southwest part of the bar there is a peak seen at the shorter infrared wavelengths (i.e., from  3.6 to 37\,$\mu$m), and at 160\,$\mu$m the peak switches to the northwest part of the bar (Figure \ref{fig:Source2}). The exception is the 20\,$\mu$m image where the source appears as a point source (at the southwest location) but there is little to no discernible extended emission (above the extended nebular emission of the G\ion{H}{2} region) from the rest of the bar. Furthermore, at 37\,$\mu$m the bar appears to have a slight deficit in emission between the northeast to southwest peaks, giving the impression of a double source. This source was not covered by the \citet{2003AA...400..223N} mid-infrared maps, so we have no higher resolution observations two confirm or deny this claim. Therefore, there could be a younger and/or more embedded source in the northeast (as evidenced by the rising SED into the far-IR), and an older and/or less embedded source to the southwest (as evidenced by the bright peak seen in the near-IR), which are not completely resolved from each other and together give the impression of bar of emission. Alternatively, the southwest peak could be the only YSO and the extended emission is the envelope from which it formed. Because we cannot clearly resolve two sources in the multi-wavelength data, we modeled this source as a single source and find it to be well-fit by a range of MYSO models from 16-32\,$M_{\sun}$ with a best-fit mass of 24\,$M_{\sun}$. Interestingly,  the radio continuum maps of \citet{1999AJ....117.2902D} show no radio emission peak at this location, and therefore if the object is indeed a MYSO, it must be a very young MYSO prior to the onset of producing significant ionizing radiation.

\subsubsection{Source 3 \& Radio Source B}
\begin{figure*}[tb!]
\epsscale{1.17}
\plotone{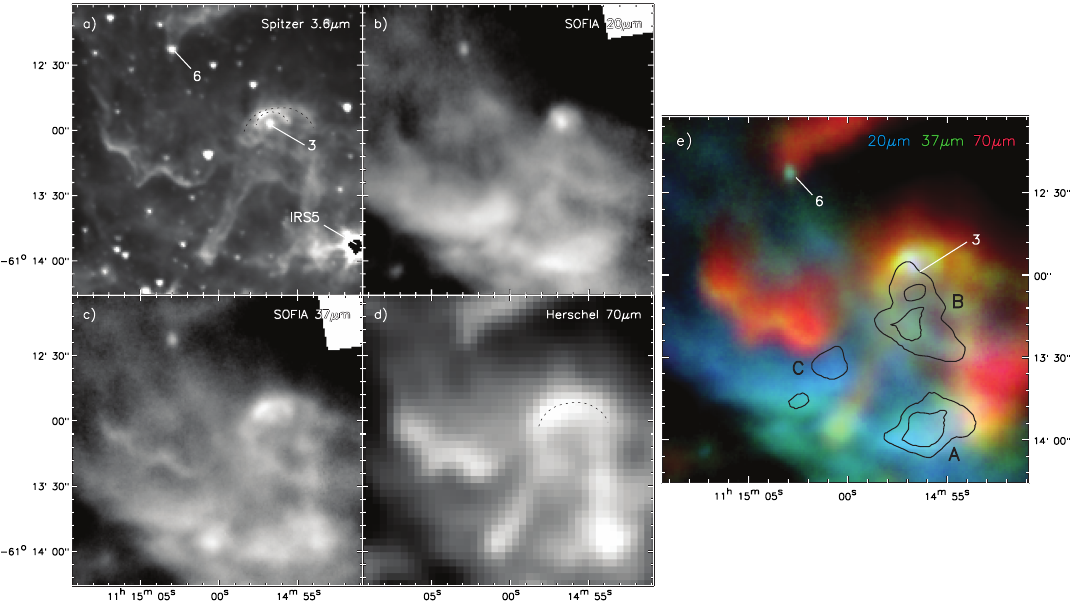}
\caption{Images of sources 3 and 6 and the radio A, B, and C regions at a) Spitzer 3.6\,$\mu$m, b) SOFIA 20\,$\mu$m, c) SOFIA 37\,$\mu$m, and d) Herschel 70\,$\mu$m. In panel e) the region is shown as a three-color infrared composite with the radio continuum contours of \citet{1999AJ....117.2902D} overlaid and the major radio regions labeled.  \label{fig:Source3}}
\end{figure*}

Radio continuum region B \citep{1999AJ....117.2902D}  is a triangular-shaped region whose northern vertex appears to be nearly coincident with  source 3. There are infrared ridge structures (best seen at 37\,$\mu$m in Figure \ref{fig:Source3}) that mimic this triangular shape, and thus it may be that it is an aspherical compact \ion{H}{2} region or ionized cavity. Source 3 itself can be seen in all Spitzer-IRAC bands as a point source with an arc of emission to the north partially surrounding it at a radius of $\sim$5$\arcsec$ (Figure \ref{fig:Source3}a). This appears to be embedded in a larger arc of emission ($\sim$30$\arcsec$ across) that can be seen well at 37\,$\mu$m and is the only structure that can be seen at 70\,$\mu$m (Figure \ref{fig:Source3}d). Interestingly, this larger arc is not seen at 20\,$\mu$m, and the point source at inner arc are unresolved from each other. In particular, at 70\,$\mu$m this arc appears to extend and almost completely surround the radio continuum radio. Given that there is no star seen in the center of the radio continuum region, it may be that point source component of source 3 is a revealed ionizing star responsible for the radio continuum emission and the asymmetry of the ionized region may be due to environmental conditions (i.e., there is denser material to the north). Contrary to this however, the CS map of \citet{2002A&A...394..253N} shows that there is a clump here called MM6 whose molecular emission appears to peak south of source 3, between radio sources A and B. This region is not covered by either the \citet{2003AA...400..223N} mid-infrared maps. Using photometry obtained by integrating over the combined area of the point source and inner arc at all wavelengths, we find that source 3 is well-fit by a range of MYSO models from 16-24\,$M_{\sun}$ with a best-fit mass of 16\,$M_{\sun}$. 

\subsubsection{Sources 4 \& 5}
\begin{figure*}[tb!]
\epsscale{1.17}
\plotone{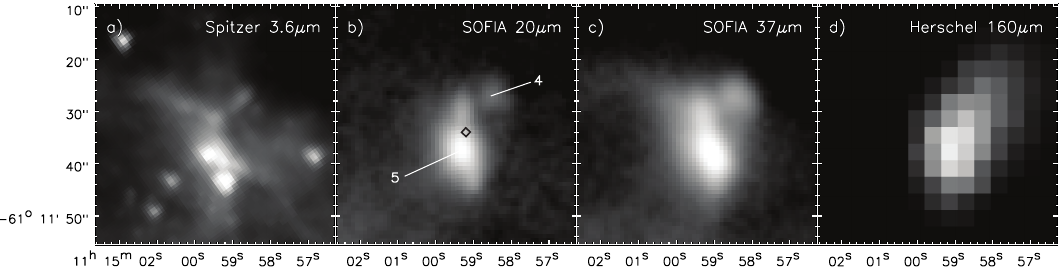}
\caption{Images of the region containing the sources 4 and 5, at a) Spitzer 3.6\,$\mu$m, b) SOFIA 20\,$\mu$m, c) SOFIA 37\,$\mu$m, and d) Herschel 160\,$\mu$m. The black diamond denotes the location of MYSO candidate Her-38 from \citet{2015ApJ...799..100D}. \label{fig:Source4}}
\end{figure*}

Sources 4 and 5 are the northernmost compact IR sources on our SOFIA maps. At both 20 and 37\,$\mu$m source 5 appears as to be elongated N-S, and source 4 appears as a point-like source 12$\arcsec$ to the northwest of the peak of source 5. At shorter infrared wavelengths, the two sources appear to be connected via a ridge of dust emission, and source 5 appears to have a tail toward the northeast (this tail is also seen at 37\,$\mu$m). Together the more compact source plus the extended emission appear as a `Y'-shaped morphology in the 3.6--8.0\,$\mu$m Spitzer images (Figure~\ref{fig:Source4}). However, the Spitzer images also reveal that source 5 has two near-infrared peaks, the northern one coincident with the 20 and 37\,$\mu$m peak. We conjecture that the southern near infrared peak is a more revealed stellar source as it fades rapidly towards longer wavelengths (i.e., it is completely absent in the 20\,$\mu$m image), whereas the northern near-infrared peak of source 5 is more embedded as it continues to be the peak in emission all the way out to 160\,$\mu$m. A MYSO (Her-38) is identified near here by \citet{2015ApJ...799..100D} with coordinates $\sim$3$\arcsec$ from the peak at 37\,$\mu$m, but looking at the region in the Herschel 160\,$\mu$m filter (Figure~\ref{fig:Source4}), it is likely the same source. Our SED fitting show source 5 to be best fit by a MYSO model with a mass of 16\,$M_{\sun}$ with a range of MYSO models fits from 16-32\,$M_{\sun}$. Source 4 is very faint in the Spitzer images, but becomes substantially bright enough by 160\,$\mu$m to be seen as a bright tongue of emission northwest of the peak of source 5. Our SED fitting show source 4 to be a less-massive YSO than source 5, with all well-fit MYSO models having a mass of 8\,$M_{\sun}$. Whether or not these MYSOs display any radio continuum emission is unknown, as this region is not covered by the \citet{1999AJ....117.2902D} radio continuum maps (incidentally, nor was it covered by the \citealt{2003AA...400..223N} mid-infrared maps).

\subsubsection{Source 6}
Source 6 appears as a point-like source at 20 and 37\,$\mu$m (Figure \ref{fig:Source3}). However, in the Spitzer images it can be seen that this mid-infrared peak corresponds to the location of a near-infrared peak that resides at the tip of a photoablated trunk (see Figure \ref{fig:Source3}a). This trunk points back towards the HD 97950, and thus this OB cluster is likely to be responsible for the erosion around this source (as it is for most trunks in the area). At 70 and 160\,$\mu$m the dust from the shaft and base of the trunk is more prominent than the dust near the stellar source at the tip. Our SED fitting show this source is likely to be a MYSO of 8\,$M_{\sun}$. This area is covered by the radio continuum maps of \citet{1999AJ....117.2902D}, but there is no detectable cm emission from source 6, indicating that it is likely to be a very youthful MYSO, at a stage prior to the onset of ionized emission. 

Since this MYSO is located at the tip of a photoablated trunk, one interesting consequence is that it likely will not be able to continue to accrete much more mass, even though it is a fairly youthful state, due to the fact that its reservoir of material is being stripped away.

\subsubsection{IRS4}
IRS4 was first detected in the 2\,$\mu$m images of \citet{1977ApJ...213..723F}. This is a bright stellar source that can be seen readily in the optical ($m=12$ in the GAIA G passband) but is considered to be a long-period variable candidate \citep{2020yCat.1350....0G}. It is the brightest stellar object in all of NGC\,3603 at 2MASS J-band \citep[$m_J=6.7$;][]{2003yCat.2246....0C} but it decreases in flux quickly with increased wavelength. It is saturated in all Spitzer-IRAC bands (not shown here). It is clearly detected in the higher angular resolution 12\,$\mu$m map of \citet{2003AA...400..223N}, and also appears as a prominent point source at 20\,$\mu$m (Figure~\ref{fig:fig2}), but is just barely detected at 37\,$\mu$m (Figure~\ref{fig:fig3}), and it is not detected at 70 or 160\,$\mu$m (not shown here). Given the behavior of flux as a function of wavelength alone (as well as the lack of a radio continuum component, e.g. \citealt{1999AJ....117.2902D}) it can be concluded that this is likely to be a (perhaps almost) completely revealed stellar source and the flux measured at all wavelengths is likely to be predominantly photospheric emission. Indeed, based upon the measured NIR colors \citet{1977ApJ...213..723F} claim IRS\,4 is an M supergiant. Though \citet{1977ApJ...213..723F} conjecture the source is likely to be at the distance of NGC\,3603, the GAIA parallax measurements for this source place it at 3433$\pm$28\,pc. Therefore, since NGC\,3603 is at 7.2\,kpc, this source is a foreground stellar source. Given the above information, and the fact that we only have one nominal flux data point (i.e. at 20$\mu$m), we did not attempt to use the MYSO SED fitter on this source.

\subsubsection{Source 7}
This source is overall a faint object, appearing brightest in the SOFIA 20\,$\mu$m image (Figure~\ref{fig:fig2}). It is barely visible in the Spitzer 3.6\,$\mu$m image, and it is not detected in the Herschel 70 or 160\,$\mu$m images (not shown here). It is identified as source 5A in the 12\,$\mu$m map of \citet{2003AA...400..223N}, but we do not see in the SOFIA data any hints of sources 5B or 5C\footnote{In the SOFIA data we furthermore only see the nearby components 16A and 16B, barely, at 20\,$\mu$m.}. Our SED of this source appear to turn over near 20\,$\mu$m, and the MYSO fitter does a poor job of fitting the data. We find that a simple blackbody of $\sim$220\,K fits the SED better, and therefore we believe that this source is a knot of dust and not an internally heated YSO.

\begin{figure*}[tb!]
\epsscale{1.17}
\plotone{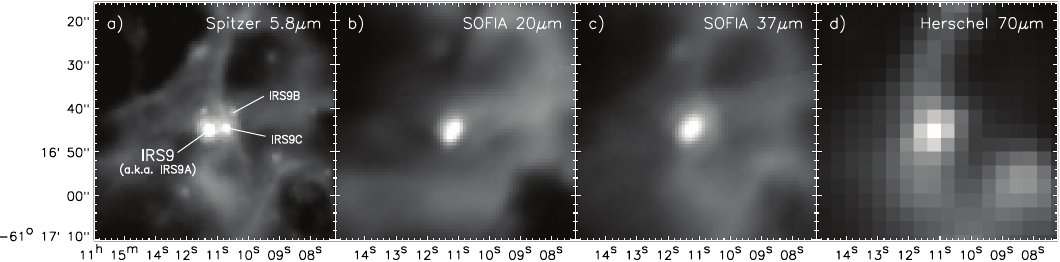}
\caption{Images of the region containing IRS9, at a) Spitzer 5.8\,$\mu$m, b) SOFIA 20\,$\mu$m, c) SOFIA 37\,$\mu$m, and d) Herschel 70\,$\mu$m. The source appears artificially elongated in the SOFIA images due to telescope issues during the observations.\label{fig:IRS9}}
\end{figure*} 

\subsubsection{Source 8}
In the Spitzer 3.6--8.0\,$\mu$m images (not shown here), this source appears as a point-like source located in (or in projection against) a large ($\sim0\farcm9$) ridge of dust running more or less east-west. A compact source is detected at 20\,$\mu$m (Figure~\ref{fig:fig2}) and at 37\,$\mu$m (Figure~\ref{fig:fig3}), however at 37\,$\mu$m there is more nearby extended nebular emission from the ridge of dust making the source appear less-prominent than at 20\,$\mu$m. In fact, the ridge of dust is not detected at 20\,$\mu$m, even though it can be readily seen at all other infrared wavelengths. Though the infrared SED of source 8 is well-fit by a range of MYSO models from 8-16\,$M_{\sun}$ with a best-fit mass of 8\,$M_{\sun}$, there appears to be no radio continuum emission present here \citep{1999AJ....117.2902D}. This means this source is likely a youthful MYSO, prior to the onset of significant ionizing emission. 

\subsubsection{Source 9}
This is another source that is compact and bright in the Spitzer-IRAC data (not shown here). It appears as an unresolved source (identified as source 11) in the 12\,$\mu$m map of \citet{2003AA...400..223N}. It is detected as a compact source at both 20\,$\mu$m (Figure~\ref{fig:fig2}) and 37\,$\mu$m (Figure~\ref{fig:fig3}), but it is not detected above the bright background of the extended nebular dust emission in the Herschel data (not shown here). Our SED fitting show it to likely be a MYSO of 12\,$M_{\sun}$. Once again, there appears to be no radio continuum emission present here \citep{1999AJ....117.2902D}, and therefore source 9 is likely to be a MYSO in the earliest stages of development.

\subsubsection{IRS 9} 
This source is the brightest compact source on the entire NGC\,3603 field at both 20 and 37\,$\mu$m (see Figure~\ref{fig:IRS9}). First detected by \citet{1977ApJ...213..723F}, it can also be seen in the 12\,$\mu$m and 18\,$\mu$m maps of \citet{2003AA...400..223N}. While the source appears elongated to the northwest in the SOFIA data, this is the direction the telescope was chopping, and this elongation is therefore not intrinsic to the source but due to problems with the stability of the chopping mechanism during these observations (see discussion in Section\,\ref{sec:obs}). There are two additional infrared point sources located west of the main IRS9 peak, as first detected in the higher angular resolution 12 and 18\,$\mu$m data of \citet{2003AA...400..223N} and \citet{2003A&A...404..255N}, and seen in the Spitzer data (Figure~\ref{fig:IRS9}a). Those authors call the main bright point source IRS9A (coincident with the SOFIA 20 and 37\,$\mu$m peak), and the fainter sources IRS\,9B and 9C. IRS9B and C are not seen in the SOFIA images (Figure~\ref{fig:IRS9}b and c), and the peak seen at both 70 and 160\,$\mu$m is coincident with the location of IRS9A (Figure~\ref{fig:IRS9}d). IRS9B and C have optical emission as seen with GAIA, as does IRS9A (DR3 5337417774995179776; $m_G \sim 17$), though no distance determination is listed in that catalog for any of the sources.

Despite the prominence of IRS9A across the entire infrared, it lacks detectable radio continuum emission. Though extended radio continuum emission does goes through the area of this source, it has no peak located here. \citet{1977ApJ...213..723F} claim that it is the reddest point source on the field and thus must be surrounded by significant dust, and therefore it is likely to reside within NGC~3603. Our SED modeling show this source is well-fit by a range of MYSO models from 32-64\,$M_{\sun}$ with a best-fit mass of 64\,$M_{\sun}$, and this is the source with the highest mass found in all of the sources we modeled. Given the lack of prominent radio continuum emission, this MYSO must also be very youthful. Given the fact that there is also a optical component seen, this may mean that the source it situated with its outflow axis pointing towards us.

\begin{figure*}[tb!]
\epsscale{1.17}
\plotone{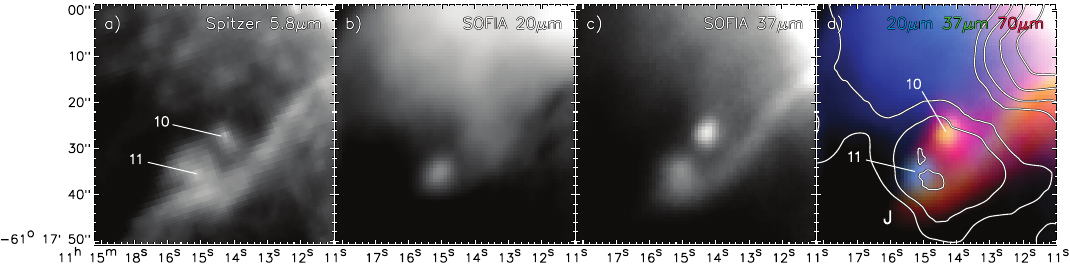}
\caption{Images of the region containing sources 10 and 11 and radio continuum peak J, with a) Spitzer 5.8\,$\mu$m, b) SOFIA 20\,$\mu$m, and c) SOFIA 37\,$\mu$m. In panel d) the region is shown as a three-color infrared composite with the radio continuum contours of source J from \citet{1999AJ....117.2902D} overlaid and labeled.  \label{fig:Source10}}
\end{figure*}

\subsubsection{Sources 10 \& 11}

Sources 10 and 11 are separated by only $\sim$11$\arcsec$ and both lie within the confines of the radio region J defined by \citet{1999AJ....117.2902D}. They both also lie just to the north of a ridge of emission, best seen in the Spitzer-IRAC bands (Figure~\ref{fig:Source10}). 

Source 10 is coincident with water maser emission \citep{1989AuJPh..42..331C}. It is barely visible in the Spitzer 3.6\,$\mu$m image but brightens at longer IRAC wavelengths, and becomes more prominent than source 11 in our 37\,$\mu$m and longer infrared wavelength data (Figure \ref{fig:Source10}). Using Herschel data, \citet{2015ApJ...799..100D}, who identify this source as Her-83, showed that the source is well fits by MYSO SED models. Interestingly, it is not seen in the 12\,$\mu$m maps of \citet{2003AA...400..223N}, and a peak is not detectable here in the SOFIA 20\,$\mu$m maps above the extended nebular emission either (Figure \ref{fig:Source10}b). Nonetheless, our SED modeling shows good fits for a stellar mass of 16\,$M_{\sun}$, but we caution that there is only one nominal data point used in the fit beyond 8\,$\mu$m (i.e. 37$\mu$m) so this result is less trustworthy.

Source 11 is coincident with a 12\,$\mu$m unresolved point source in the images of \citet{2003AA...400..223N}, which they call source 10. The mid-infrared emission can be seen at this location out to 37\,$\mu$m (Figure \ref{fig:Source10}), but is not detectable as a source at Herschel 70 and 160\,$\mu$m above the emission from the ridge of dust present here. The infrared peak of source 11 also appears to have an optical component ($m=12$ in the GAIA G passband), and in the Spitzer-IRAC 5.8 (Figure \ref{fig:Source10}a) and 8.0\,$\mu$m images, it is surrounded almost entirely by a dust shell with a radius of $\sim$4$\arcsec$. At the longer SOFIA wavelengths there is not sufficient resolution two separate the point source and shell emission, and so our photometry of this source includes an area covering both the point source and shell. 

Confusingly, the GAIA optical source found here (which has a position that agrees with the infrared peak to within 1$\arcsec$) has a distance of only 553$\pm$5\,pc. However, from the infrared fluxes of this source we find an SED that is best-fit by a MYSO model of 16\,$M_{\sun}$, with good fits ranging from 16-24\,$M_{\sun}$. Furthermore, the peak of the extended radio continuum source J peaks very close to the infrared peak of source 11 (see Figures \ref{fig:fig2} and \ref{fig:fig3}). It may be that the optical point source component and the more extended infrared component are simply a chance alignment of a field star and a MYSO, and are unrelated to each other. 

\subsubsection{Source 12}

This source is a very bright unresolved point source in the near-infrared, so much so that it is completely saturated at all Spitzer-IRAC wavelengths (not shown here). Though it is well-detected at 20\,$\mu$m (Figure~\ref{fig:fig2}), it is barely detected at 37\,$\mu$m (Figure~\ref{fig:fig3}), and not visible in either of the 70 or 160\,$\mu$m Herschel passbands (not shown here). Given that there are only two photometry points for this source (20 and 37\,$\mu$m), and that the flux decreases substantially from 20 to 37\,$\mu$m, we suggest that this source is an evolved (main sequence) star and not a YSO. Indeed, it does have an optical component, and although observed with GAIA (DR3 5337418221666129792), it is rather faint ($m_G=18.5$) and highly variable with no accurate distance derivation given in the GAIA catalog. Like IRS4, given the obvious stellar nature of this source and the lack of data points, we did not attempt to model this source with the MYSO SED fitter.

\subsubsection{Radio source F}

There is an infrared peak seen in the SOFIA 20 and 37\,$\mu$m images (see Figures~\ref{fig:fig2} and~\ref{fig:fig3}) that aligns well with the radio continuum peak labeled source F by \citet{1999AJ....117.2902D}. Radio peak F is embedded in a larger extended radio continuum emission region, and it is unclear if it is a YSO itself, or just a ionized knot of dust. Though no clear peak can be seen here at 70 or 160\,$\mu$m, there is a obvious knot of emission seen in the extended dust in this area in the Spitzer-IRAC data at all wavelengths. The SED fitting algorithm best fits this source as a 16\,$M_{\sun}$ MYSO, however the fit is poor, raising some doubt as to its true nature.

\begin{figure*}[tb!]
\epsscale{1.17}
\plotone{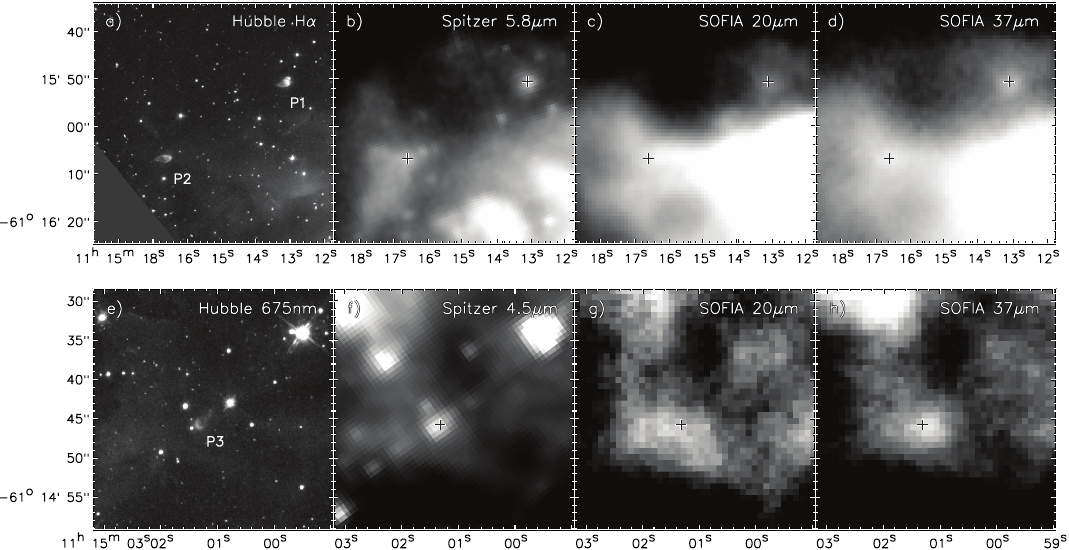}
\caption{Images of the regions containing the proplyd sources P1 and P2 (top row) and P3 (bottom row). Panels a) and e) show Hubble images, panels b) and f) show Spitzer-IRAC images, panels c) and g) show the SOFIA 20\,$\mu$m images, and panels d) and g) show the SOFIA 37\,$\mu$m images. The proplyds are labeled in the Hubble images and their positions are given by the crosses in all other images. SOFIA images were shifted by one pixel (0.768$\arcsec$) in R.A. to better align to the Hubble data. \vspace{0.3in} \label{fig:Proplyds}}
\end{figure*} 

\subsubsection{The Proplyds: P1, P2, and P3}

Based upon HST/WFPC2 observations in the optical and VLT/ISAAC observations in the near-infrared, \citet{2000AJ....119..292B} discovered three tadpole or cometary shaped objects within NGC\,3603, possessing ionization fronts at their heads, which face the HD 97590 stellar cluster, and ionized tails pointing away from the cluster. It is claimed by \citet{2000AJ....119..292B} that these structures are analogous to the proplyds (PROto PLanetarY DiskS) seen in Orion \citep{1993ApJ...410..696O}, though they are 20 to 30 times larger in size.  

There exists some diffuse 12\,$\mu$m emission in vicinity of P1 as seen by \citet{2003AA...400..223N}. In our SOFIA 20 and 37\,$\mu$m images we definitively see a peak a the location of P1 at both wavelengths (Figure \ref{fig:Proplyds}c \& d), and it is clearly detected at all Spitzer-IRAC bands (e.g., Figure \ref{fig:Proplyds}b) . Even in the Herschel 70 and 160\,$\mu$m images there appears to be a tongue of faint emission extending out (but only partially resolved) from  the bright infrared emission region associated with radio source E. Attempting to do the best we could to isolate emission just from P1, we performed SED fitting to the photometry of the source and find that it is well-fit by several MYSO models all having a mass of 8\,$M_{\sun}$.

Though there is a weak peak seen at the location of P2 in all infrared wavelengths from 3.6-160\,$\mu$m, as \citet{2003AA...400..223N} point out, at most of these wavelengths the emission from a ridge of dust extended through here that makes it difficult to isolate the flux from just the proplyd itself at most wavelengths. The exception to this is the 70\,$\mu$m Herschel images, where P2 seems to be a point-like source at this wavelengths. Again, we isolated the emission from just P2 at all wavelengths as best as we could, and we were able to use the SED fitter to fit the data with a range of MYSO models from 12-32\,$M_{\sun}$ with a best-fit mass of 12\,$M_{\sun}$.

Proplyd P3 is clearly seen at all Spitzer-IRAC wavelengths (e.g., Figure \ref{fig:Proplyds}f, though at 8\,$\mu$m there are significant array artifacts that make the data unusable for photometry). The higher angular 12\,$\mu$m images of \citet{2003AA...400..223N} show a faint detection of emission at the location of P3. In the SOFIA data there exists a diffuse region of emission at this location with a broad peak near P3 at 20$\mu$m (Figure \ref{fig:Proplyds}g), and a more definitive detection of a peak at 37$\mu$m (Figure \ref{fig:Proplyds}h). At 70 and 160\,$\mu$m, there appears to be a peak here as well, but like the SOFIA data, it is not fully resolved from nearby extended emission features. Again, isolating the flux from just P3 we found the fluxes were well-fit by a range of MYSO models from 8-16\,$M_{\sun}$ with a best-fit mass of 12\,$M_{\sun}$, however, this fit is only based upon 2 nominal data points (and 5 upper limits).

\begin{figure*}[tb!]
\epsscale{1.00}
\plotone{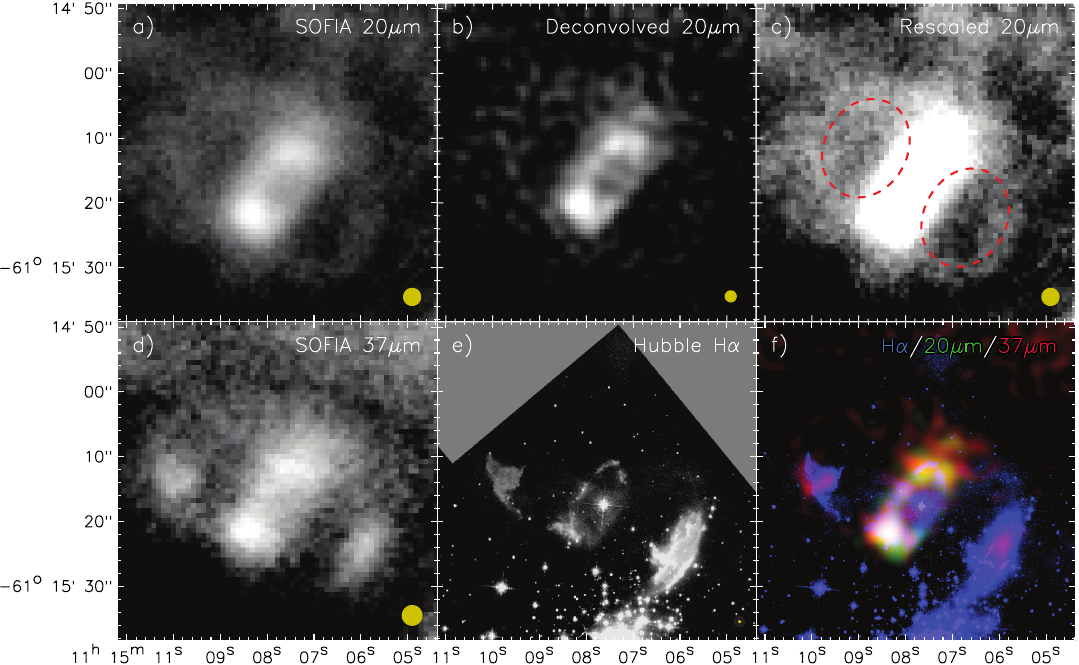}
\caption{Images of circumstellar ring and hourglass nebula of Sher\,25. a) The 20\,$\mu$m by SOFIA. b) The deconvolved 20\,$\mu$m SOFIA image showing the clumpy dust structures within the circumstellar ring. c) The 20\,$\mu$m by SOFIA rescaled to show the fainter emission from the hourglass nebula (traced by the dotted red lines). d) The 37\,$\mu$m by SOFIA. e) The Hubble H$\alpha$ image. f) A 3-color image made from the Hubble H$\alpha$ image (blue), the SOFIA deconvolved 20\,$\mu$m image (green), and the SOFIA deconvolved 37\,$\mu$m image (red). The yellow filled circles in the lower right corner of panels a-e indicate the resolution of the image at the given wavelength.\label{fig:Sher25}}
\end{figure*}

At a minimum, the detection of infrared emission at thermal infrared wavelengths for all three sources (though varied in their strength), indicates that these sources do indeed possess circumstellar dust of some kind. Our SED modeling shows that these sources may indeed be legitimate MYSOs, and therefore may possess not only circumstellar disks (like proplyds), but they are likely to also be surrounded by dust envelopes as well. This results seems to contradict the speculations by \citet{2003AA...400..223N} that these sources may be small, denser clumps of gas that have simply not yet been photo-evaporated away by the OB stars present in NGC\,3603, constituting the last remnants of environmental over-densities present in the original molecular cloud. Furthermore, proplyds can be confused with cometary-shaped UC\ion{H}{2} regions, with the main distinction being that proplyds are externally ionized, whereas UC\ion{H}{2} regions have a central massive star responsible for ionizing the structure internally. The results of the radio study by \citet{2002ApJ...571..366M} showed that the radio flux of the proplyds in NGC\,3603 could be entirely attributed to external ionization by the HD\,97590 stellar cluster alone, and they speculate that it is unlikely that any central stars to these sources would be $>10M_{\sun}$. However, our conclusions are that these objects may indeed contain YSOs massive enough to ionize their surroundings, and thus it is unclear if these are genuine proplypds or cometary UC\ion{H}{2} regions.    

\subsubsection{Sher 25}

Located about 20$\arcsec$ north of the HD\,97950 star cluster, Sher\,25 is an evolved blue supergiant (BSG), discovered by \citet{1965MNRAS.129..237S} at optical wavelengths and later spectral typed by \citet{1983A&A...124..273M} to be a B1.5\,Iab star. \citet{1997ApJ...475L..45B} discovered the BSG is surrounded by a circumstellar ring with a diameter of $\sim$0.4\,pc and detect what they believe are bipolar outflow clouds located northeast and southwest on the ring separated $\sim$20\,pc from the central star.  

The clumpy circumstellar ring structure and bipolar clouds are best seen in the HST images of \citet{2000AJ....119..292B}, especially in the H$\alpha$ filter (Figure\,\ref{fig:Sher25}e). Our SOFIA 20\,$\mu$m map shows dust emission present around the entire circumstellar ring, and also shows faint emission from the East Cloud (to use the nomenclature of \citealt{2008MNRAS.388.1127H}) of the bipolar clouds (Figure\,\ref{fig:Sher25}a). The emission from the circumstellar ring is enhanced substantially in the northwest and southeast ends of the ring, which would be expected for a ring tilted to our line of sight (i.e., 64$\arcdeg$ against the plane of the sky, and with a position angle of $\sim$165$\arcdeg$, as measured by \citealt{1997ApJ...475L..45B}) as the pathlength of dust (and hence quantity of emitting material) is larger at these locations. That being said, as seen is Figure\,\ref{fig:Sher25}b, our deconvolved 20\,$\mu$m SOFIA image shows clumpy structures co-located with the clumps seen in the Hubble H$\alpha$ image (Figure\,\ref{fig:Sher25}e). The SOFIA 37\,$\mu$m image shows dust emission mostly from the northwest and southeast ends of the ring, but both the East and West Clouds are well detected (Figure\,\ref{fig:Sher25}d). Figure\,\ref{fig:Sher25}f shows the good spatial correlation between the structures seen in the SOFIA images and the Hubble H$\alpha$ images. In the circumstellar ring the SOFIA clumps are seen to be co-located with the H$\alpha$ clumps, and the emission at 37\,$\mu$m form the East and West Clouds can be seen to align well with the structures seen in the Hubble image as well.

\begin{deluxetable*}{lcclc}
\tabletypesize{\scriptsize}
\tablecolumns{5}
\tablewidth{0pt}
\tablecaption{Updated Coordinates and Distances of the Near-Infrared IRS Sources in NGC 3603}\label{tb:IRS}
\tablehead{
           \colhead{ Source }&
           \colhead{ R.A.}&
           \colhead{ Decl. }&
           \colhead{ Dist.$^\ddagger$}&
           \colhead{ GAIA DR3 ID}\\
	   \colhead{  } &
           \colhead{(J2000) }&
           \colhead{(J2000) }&
           \colhead{ (pc)}&
           \colhead{ }
}
\startdata
IRS4 	&11:15:03.62 		&-61:14:22.3 &3433$_{-28}^{+28}$   &5337418462189917952\\
IRS5	&11:14:51.31		&-61:13:52.6 &4914$_{-30}^{+12}$   &5337421043475869184\\
IRS6 	&11:15:25.91		&-61:13:41.8 &8022$_{-936}^{+378}$ &5337419772188803712\\
IRS7	&11:14:59.25		&-61:19:28.4 &5833$_{-48}^{+70}$   &5337042399168099072\\
IRS8$^\dagger$	&11:15:08.93		&-61:16:00.4 &4274$_{-94}^{+54}$   &5337418015513337472\\
IRS9	&11:15:11.38		&-61:16:45.2 &\nodata   &5337417774995179776\\
IRS12	&11:14:58.28		&-61:17:17.5 &2866$_{-111}^{+91}$   &5337417912434087680\\
IRS13	&11:14:54.00		&-61:18:35.4 &\nodata   &5337042570966803328\\
IRS15   &11:15:25.24   		&-61:12:01.5 &\nodata   &5337419905291591296
\enddata
\tablecomments{These are updated coordinates for the near-infrared sources given by \citet{1977ApJ...213..723F}. New centroid data is from the closest bright source seen in the 2MASS 2\,$\mu$m data. No clear near-infrared source can be found for IRS14, and there are no IRS10 or 11 coordinates tabulated in \citet{1977ApJ...213..723F} as they were too faint.}
\tablenotetext{\dagger}{Distances are from the GAIA Data Release 3 (DR3) catalog and determined from parallax measurements. The distances to IRS9, IRS13, and IRS15 are not given in the catalog.}
\tablenotetext{\dagger}{IRS8 has two sources within the original $\sim$10$\arcsec$ source position error. The coordinates and information are for the brightest of the two sources. The fainter source is GAIA DR3 5337418015513340160 and is located 5$\arcsec$ southwest of this source and has no GAIA-determined distance.}
\end{deluxetable*}

The only dust emission seen in the Spitzer 3.6\,$\mu$m image is from the East Cloud (and the BSG is also apparent at this wavelength as well as all other Spitzer-IRAC wavelengths). Emission from both clouds and the circumstellar ring are evident in the Spitzer images at 4.5, 5.8, and 8.0\,$\mu$m. At Herschel 70\,$\mu$m, emission can be seen coming from all three objects as well, but at 160\,$\mu$m the Clouds are more apparent than emission from the ring. From this behaviour of flux as a function of wavelength alone, it can be ascertained that the dust emission in the Clouds is cooler than the dust in the ring. 

It was hypothesized by \citet{1997ApJ...489L.153B} that the East and West Clouds are part of a hourglass-shaped circumstellar nebula, and that the hourglass nebula and ring are akin to similar structures seen around SN\,1987A,  which were believed to have been created by its supergiant progenitor star prior to going supernova. Hubble images by \citet{1995ApJ...452..680B} showed that the hourglass-shaped nebula around SN\,1987A is best seen in narrow-band H$\alpha$ images. Such a hourglass-shaped nebula is not readily apparent in the Hubble H$\alpha$ image in Figure\,\ref{fig:Sher25}e, however \citet{1997ApJ...489L.153B} showed that there is some (less-compelling) evidence in their Hubble near-infrared spectroscopic channel maps which show a hint of a broken outline of an hourglass structure seen in the H$\alpha$ line. Perhaps lending credence to the existence of this structure, our SOFIA image at 20\,$\mu$m shows emission outlining the entire hourglass-shaped nebula (see red dashed ellipses in Figure\,\ref{fig:Sher25}c), though the northeastern lobe appears to be filled in more with infrared emission than the southwestern lobe.    

Sher\,25 is optically bright enough that it is included in the GAIA DR3 catalog. Based upon parallax measurements, Sher\,25 is located at 5873$_{-102}^{+30}$\,pc, placing it in the foreground of NGC\,3603 by more than a kiloparsec. As the distance to NGC\,3603 was also determined using GAIA (DR2) parallaxes, this means that Sher\,25 is unlikely to be directly associated with the star-forming activity of NGC\,3603. A similar conclusion was recently made by \citet{2023arXiv230806164W} using distances derived from both GAIA data and spectrophotometric techniques.

\subsubsection{IRS sources from Frogel et al. (1977)}

The coordinates of IRS\,1 and 2 define the peaks seen in the low-resolution 10 and 20\,$\mu$m images of \citet{1977ApJ...213..723F}. These correspond to the brightest mid-infrared emission we see in the SOFIA data centered roughly on the E and H radio peaks, respectively. IRS3 also is defined in that same paper by a structure in the 10\,$\mu$m image, which appears as a tongue of emission towards the west of IRS\,1. While there is extended emission present here at all higher resolution near-infrared, as well as mid-infrared and far-infrared images, there is nothing that could be identified as a peak or separate region (Figures~\ref{fig:fig2} and~\ref{fig:fig3}). 

The rest of the IRS sources (i.e., IRS\,4-9 and 12-15; there are no IRS\,10 or 11) are defined by the 2\,$\mu$m data of \citet{1977ApJ...213..723F}. Of these near-infrared-defined sources, only IRS\,14 does not have an obvious source in the Spitzer data or 2MASS data, while the rest have clearly visible and bright Spitzer-IRAC sources (usually saturated) and 2MASS 2\,$\mu$m sources to within the quoted 10$\arcsec$ astrometric accuracy. In our figures in this paper (best seen in Figures~\ref{fig:fig2} and~\ref{fig:fig3}), we have identified and labeled the locations of the 2MASS sources we believe to be associated with each IRS source, and have tabulated the centroids based upon the 2MASS 2\,$\mu$m data in Table \ref{tb:IRS}. IRS\,4 and IRS\,9 are the only near-infrared-defined sources with emission peaks seen in our SOFIA data.

We have also searched for each of these sources in the GAIA DR3 catalog, and indeed all are found to have an optical counterpart within 1$\arcsec$ of their near-infrared peak positions (i.e., coincident to within our infrared positional measurement uncertainty). The DR3 identification numbers for each IRS source are given in Table\,\ref{tb:IRS}, along with their distances derived from their measured GAIA parallaxes, as given in the DR3 catalog. IRS9, IRS13, and IRS15 do have optical components, but accurate distances were unavailable in the catalog.   

It can be seen from the distances in Table\,\ref{tb:IRS} that most of the IRS sources with GAIA distances are not physically associated with NGC\,3603. Most appear to be foreground stars, with the exception of IRS6. At 8022$_{-936}^{+378}$\,pc, this source lies at the same distance as NGC\,3603 ($\sim$7.2\,kpc) to within the errors.  

\section{Results and Data Analysis}\label{sec:data}

We subdivide the resolved sources of infrared emission within NGC~3603 into compact or extended sub-regions. The two categories denote which objects we believe are star-forming cores (compact sources) vs. the larger star-forming molecular clumps (extended sub-regions). For the compact sources, we will apply SED models to their multi-wavelength photometry to estimate their physical characteristics and to determine which sources are likely to be MYSOs. For the extended sub-regions, we will estimate their relative evolutionary states using analyses based upon their derived infrared mass, luminosity, and gas kinematics to discern information about the evolutionary state of NGC~3603. 

\subsection{Physical Properties of Compact Sources: SED Model Fitting and Determining MYSO Candidates}\label{sec:cps}

We define a compact source as one that has a definitive peak that does not change location significantly with wavelength, and it must also be detected at more than one wavelength. As was the case in our previous papers, the compact sources chosen have physical sizes $\lesssim0.3$\,pc, which is consistent with the size of molecular cores which are on the order of $\sim0.1$ pc \citep{2007ARA&A..45..481Z}. We measured flux densities for all compact sources and sub-regions that could be identified in the SOFIA 20\,$\mu$m and 37\,$\mu$m data. We additionally downloaded Spitzer-IRAC (i.e., 3.6, 4.5, 5.8, 8.0\,$\mu$m) imaging data and Herschel-PACS (i.e., 70 and 160\,$\mu$m) imaging data from their respective online archives and measured fluxes for these same sources at all wavelengths.  Table~\ref{tb:SOFIA_compact} contains the information regarding the position, radius employed for aperture photometry, and background subtracted flux densities measured at both SOFIA wavelengths for all compact sources (and similar information regarding photometry from the Spitzer and Herschel data can be found in Appendix~\ref{sec:appendixflux}). We employed the same optimal extraction technique as in \citetalias{2019ApJ...873...51L} to find the optimal aperture to use for photometry. Background subtraction was also performed in the same way as \citetalias{2019ApJ...873...51L} (i.e. using background statistics from an annulus outside the optimal extraction radius which had the least environmental contamination).

We found eighteen compact sources in the SOFIA data, and eight are newly identified here. Using the cm radio continuum maps of \citet{1999AJ....117.2902D}, and archival ATCA 6\,cm data that are available at similar resolution but with slightly different field coverage, we find that most of these infrared compact sources (14 of 18) do not correspond to cm radio continuum peaks or compact sources. Three of the compact sources are the previously identified proplyds, which are known to have cm radio continuum emission from the study of \citet{2002ApJ...571..366M}. The only other compact infrared object with detected cm radio continuum emission is radio source F \citep{1999AJ....117.2902D}. However, most of the infrared compact sources are found to lie embedded within areas of diffuse and extended radio continuum emission, with the exception of sources 6, 7, 8, and IRS4 where there is no coincident or environmental radio emission detected at all. Sources 4 and 5 are not covered by the \citet{1999AJ....117.2902D}  or archival 6\,cm maps and so whether or not they have associated radio continuum emission peaks is unknown.  

Most radio-defined sources from \citet{1999AJ....117.2902D} are found to be associated with mid-infrared peaks or with enhanced mid-infrared emission regions. The exceptions are radio sources K, L, and M which lie within extended and diffuse areas of infrared emission but are not associated with any infrared peak or particularly bright infrared regions.

\begin{figure}
\center
\begin{tabular}[b]{c@{\hspace{-0.1in}}c}
\includegraphics[width=3.2in]{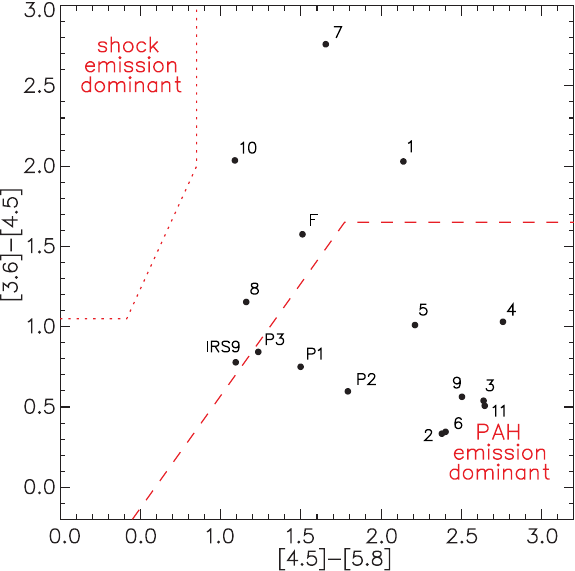}\\
\end{tabular}
\caption{\footnotesize A color-color diagram utilizing our background-subtracted Spitzer-IRAC 3.6, 4.5, and 5.8\,$\mu$m source photometry to distinguish ``shocked emission dominant'' and ``PAH emission dominant'' YSO candidates from our list of sub-components and point sources. Above (up-left) of dotted line indicates shock emission dominant regime. Below (bottom-right) dashed line indicates PAH dominant regime. We adopt this metric from \citet{2009ApJS..184...18G}. Some sources are not included in this diagram due to non-detection or saturation in the Spitzer-IRAC bands.}\label{fig:ccd}
\end{figure}

Using the SOFIA, Spitzer, and Herschel photometry, SEDs were constructed for all compact sources, except IRS4 and source 12. These two sources are saturated in all of the Spitzer wavelengths and are not detected in the Herschel filters, and thus only have two nominal flux values (at 20 and 37\,$\mu$m). As discussed in \citetalias{2019ApJ...873...51L}, a color-color diagram using Spitzer-IRAC data (3.6\,$\mu$m-4.5\,$\mu$m vs. 4.5\,$\mu$m-5.8\,$\mu$m) can be used to determine if sources are highly contaminated by shock emission and/or PAH emission, and we have employed that technique here again for the compact sources in NGC~3603. From Figure \ref{fig:ccd} we see that there are no sources classified as ``shock emission dominated'', however there are a fair number of ``PAH emission dominated'' sources. For the sources in this latter category, their 3.6, 5.8, and 8.0\,$\mu$m IRAC fluxes are set as upper limits to the photometry used in constructing the SEDs. Additionally, the Herschel 70 and 160\,$\mu$m fluxes are set to be upper limits in the SEDs for most sources due to the coarser spatial resolution ($\sim$10$\arcsec$) of the data and the high likelihood that the photometry is contaminated by emission from adjacent sources or the extended dusty environment of NGC~3603.

As we did in \citetalias{2019ApJ...873...51L}, we set the upper error bar on our photometry as the subtracted background flux value (since background subtraction can be highly variable but never larger than the amount being subtracted), and the lower error bar values for all sources come from the average total photometric error at each wavelength (as discussed in Section 2 and \citetalias{2019ApJ...873...51L}) which are set to be the estimated photometric errors of 20\%, 15\%, and 10\% for 4.5, 20, and 37\,$\mu$m bands, respectively. We assume that the photometric errors of the Spitzer-IRAC 3.6, 5.8, and 8.0\,$\mu$m fluxes are 20\% for the sources that are not contaminated by PAH features. Also as in \citetalias{2019ApJ...873...51L}, the error bars of the Herschel 70 and 160\,$\mu$m data points are assumed to be 40\% and 30\%, respectively.

Once SEDs were constructed from the photometric data (and their associated errors or limits), we utilized the ZT \citep{2011ApJ...733...55Z} MYSO SED model fitter as we did in \citetalias{2019ApJ...873...51L} in order to investigate the physical properties of individual sources. Perhaps more well-known is the SED fitter by \citet{2007ApJS..169..328R}, but their YSO models were developed mostly with the intention of fitting lower-mass protostars that are typically observed in lower pressure environments and with lower accretion rates than the massive protostars the ZT models were developed for. Some comparisons between the models fit to MYSO SEDs by both the ZT and Robitaille algorithms were given in \citet{2017ApJ...843...33D}, and while some model parameters were shown to vary significantly between the two, the values for central stellar mass and bolometric luminosity (i.e., those parameters we are concerned with more here) were in fairly good agreement.

The ZT fitter pursues a $\chi^2$-minimization to determine the best fit MYSO models, with each model fit providing a normalized minimum $\chi^2$ value (so called $\chi_{\rm nonlimit}^2$). To be consistent with the analysis of \citetalias{2019ApJ...873...51L}, we selected a group of models that show $\chi_{\rm nonlimit}^2$ values similar to the best fit model and distinguishable from the next group of models showing significantly larger $\chi_{\rm nonlimit}^2$ values. Typically the jump in $\chi_{\rm nonlimit}^2$ value from one group to the next is a factor of 3 or more than the average separation of $\chi_{\rm nonlimit}^2$ values in the preceding group. Sometimes the first and/or second best fits have significantly lower $\chi_{\rm nonlimit}^2$ values than those that come in the grouping after, and in such cases we will include those first fits with the first grouping so that we a always have at least 5 best-fit models.  

\begin{figure*}[tb!]
\epsscale{1.00}
\plotone{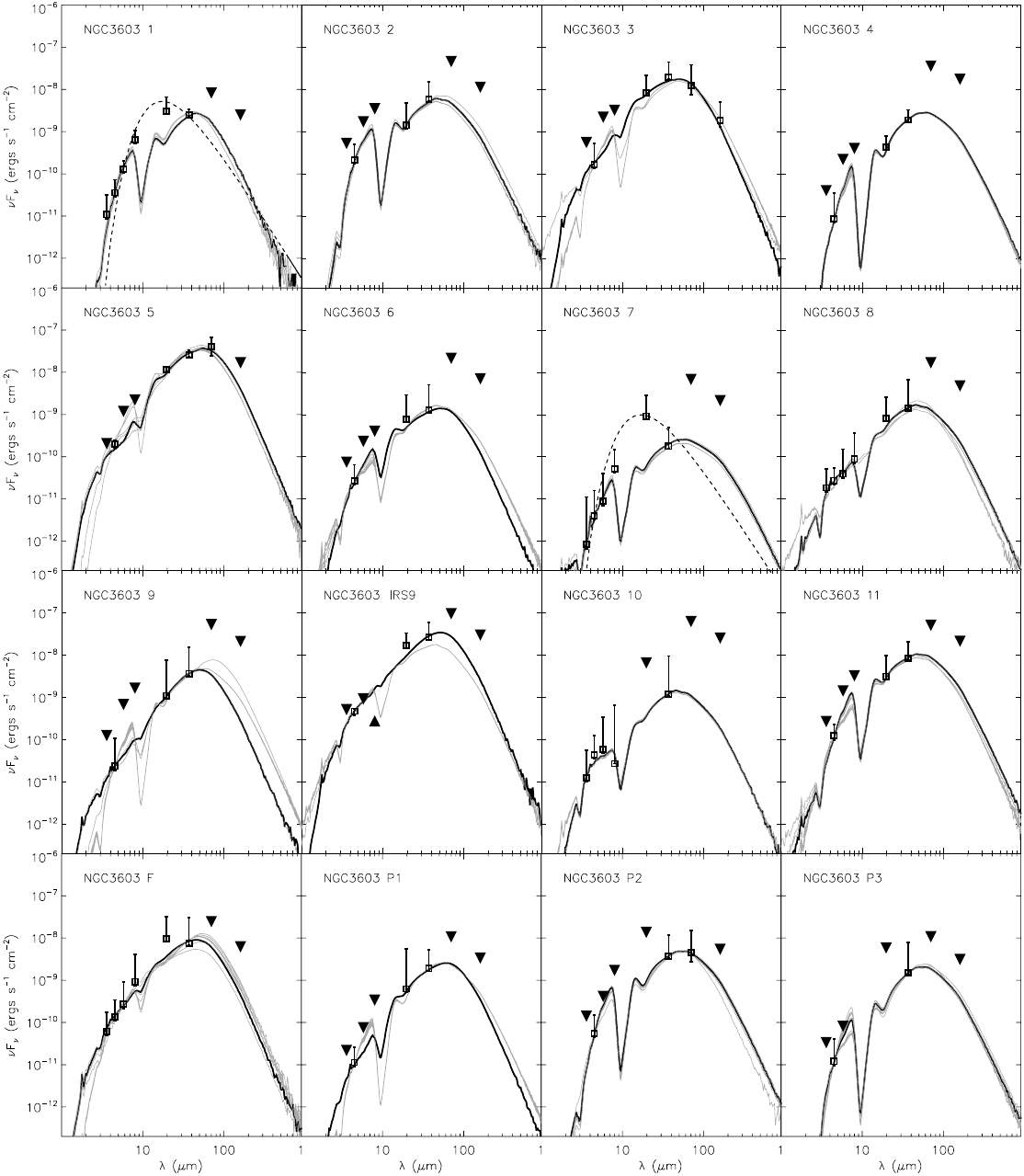}
\caption{SED fitting with ZT model for compact sources in NGC\,3603. Black lines are the best fit model to the SEDs,and the system of gray lines are the remaining fits in the group of best fits (from Table \ref{tb:sed}).  Upside-down triangles are data that are used as upper limits in the SED fits.\label{fig:SED}}
\end{figure*}

\begin{deluxetable*}{rccccrclrclcl}
\tabletypesize{\small}
\tablecolumns{12}
\tablewidth{0pt}
\tablecaption{SED Fitting Parameters of Selected Compact Infrared Sources in NGC\,3603}\label{tb:sed}
\tablehead{\colhead{   Source   }                                              &
           \colhead{  $L_{\rm obs}$   } &
           \colhead{  $L_{\rm tot}$   } &
           \colhead{ $A_v$ } &
           \colhead{  $M_{\rm star}$  } &
           \multicolumn{3}{c}{$A_v$ Range}&
           \multicolumn{3}{c}{$M_{\rm star}$ Range}&
           \colhead{ Best }&
           \colhead{Notes}\\
	   \colhead{        } &
	   \colhead{ ($\times 10^3 L_{\sun}$) } &
	   \colhead{ ($\times 10^3 L_{\sun}$) } &
	   \colhead{ (mag.) } &
	   \colhead{ ($M_{\sun}$) } &
       \multicolumn{3}{c}{(mag.)}&
       \multicolumn{3}{c}{($M_{\sun}$)}&
       \colhead{  Models   } &
       \colhead{   }
}
\startdata
1  &      5.47 &    457.90 &     79.5 &     48.0 &  79.5 & - &  79.5 &  16.0 & - &  48.0 &  7 & dust shell\\ 
2  &     14.15 &     78.27 &     62.9 &     24.0 &  57.0 & - &  74.6 &  16.0 & - &  32.0 &  6 & MYSO\\
3  &     43.32 &    112.66 &     26.5 &     16.0 &   7.9 & - &  26.5 &  16.0 & - &  24.0 &  7 & MYSO\\
4  &      6.51 &      9.45 &     31.9 &      8.0 &  28.5 & - &  39.4 &   8.0 & - &   8.0 & 13 & MYSO\\
5  &     82.72 &    115.52 &     26.5 &     16.0 &  26.5 & - &  79.5 &  16.0 & - &  32.0 &  6 & MYSO\\
6  &      3.41 &     13.30 &     53.0 &      8.0 &   8.4 & - &  53.0 &   8.0 & - &   8.0 & 10 & MYSO\\
7  &      0.68 &      0.89 &      3.4 &      4.0 &   0.8 & - &  13.4 &   4.0 & - &   4.0 &  5 & dust clump\\ 
8  &      3.90 &     10.18 &      5.9 &      8.0 &   5.9 & - &  10.9 &   8.0 & - &  16.0 & 10 & MYSO\\
9 &      9.22 &     49.40 &     26.5 &     12.0 &  22.6 & - &  53.0 &   8.0 & - &  12.0 & 12 & MYSO\\
10 &      3.21 &     28.78 &     34.4 &     16.0 &   9.2 & - &  36.9 &   8.0 & - &  16.0 &  8 & MYSO\\
11 &     25.63 &     36.53 &      8.4 &     16.0 &   1.7 & - &  31.9 &  16.0 & - &  24.0 & 11 & MYSO\\
IRS9 &     76.15 &    976.81 &     26.5 &     64.0 &   1.7 & - &  26.5 &  32.0 & - &  64.0 &  6 & MYSO\\
F  &     20.91 &    108.08 &     26.5 &     16.0 &  25.2 & - &  53.0 &  12.0 & - &  64.0 &  7 & MYSO\\
P1 &      5.75 &     15.34 &     53.0 &      8.0 &  29.3 & - &  53.0 &   8.0 & - &   8.0 & 10 & MYSO\\
P2 &     11.75 &     19.87 &     38.6 &     12.0 &  38.6 & - &  83.8 &  12.0 & - &  32.0 &  6 & MYSO\\
P3 &      4.59 &     17.02 &     67.1 &     12.0 &  51.1 & - & 108.6 &   8.0 & - &  16.0 & 11 & MYSO
\enddata
\tablecomments{\footnotesize  A ``MYSO'' in the right column denotes a MYSO candidate. Compact sources IRS4 and source 12 are not tabulated due to lack of data points for the SED fitter.}
\end{deluxetable*}

Figure~\ref{fig:SED} shows the ZT MYSO SED model fits as the solid lines (black for the best model fit, and gray for the rest in the group of best fit models) on top of the derived photometry points for each individual source. Table~\ref{tb:sed} lists the physical properties of the MYSO SED model fits for each source. The observed bolometric luminosities, $L_{\rm obs}$, of the best fit models are presented in column~2 and the true total bolometric luminosities, $L_{\rm tot}$ (i.e. corrected for the foreground extinction and outflow viewing angles), in column~3. The extinction and the stellar mass of the best models are listed in column~4 and 5, respectively. In column~6, we provide the number of the models in the group of best model fits. Columns~7 and 8 present the ranges of the foreground extinction and stellar masses derived from the models in the group of best model fits in column~6. Column~9 shows the identification of the individual sources based on the previous studies as well as our criteria of MYSOs and possible MYSOs (``pMYSOs'') set in \citetalias{2019ApJ...873...51L}. To summarize, the conditions for a source to be considered a MYSO is that it must 1) have an SED reasonably fit by the MYSO models, 2) have a M$_{\rm star}\ge8\,$M$_{\sun}$ for the best model fit model, and 3) have M$_{\rm star}\ge8\,$M$_{\sun}$ for the range of $M_{\rm star}$ of the group of best fit models. A pMYSO fulfills only the first two of these criteria.

We find that the multi-wavelength photometry for 14 of the 16 compact sources with sufficient data sampling to create SEDs can be well fit by the MYSO models. Two sources, 1 and 7, appear to peak at 20\,$\mu$m and are better fit with single temperature blackbodies (205\,K and 220\,K, respectively; see Figure~\ref{fig:SED}), and thus are thought to maybe be externally heated dust shells or knots. Source F also has a slightly high 20\,$\mu$m flux for the MYSO SED fitter, but it is less pronounced than for sources 1 and 7. Slightly high 20\,$\mu$m fluxes have been seen for one or two compact sources in almost every G\ion{H}{2} region studied in this survey, and there is the potential that some spectral feature might be enhancing emission in this filter bandpass. In the second paper of this series (Lim et al. 2020; hereafter ``\citetalias{2020ApJ...888...98L}''), we discuss that the [\ion{S}{3}] at 18.71\,$\mu$m could potentially be bright enough to affect the measured emission in this filter for some sources or regions. In any case, for source F there are many nominal data points that are fit by the MYSO fitter, and we know that source F is a substantial radio continuum source and is indeed likely to be a MYSO (as the fits suggest). We note that no sources fulfill the ``pMYSO'' criteria in NGC~3603. 


Of the 14 sources believed to be MYSOs based on the SED fitting, only 4 (F, P1, P2, and P3) have detected cm radio continuum emission. As mention previously, sources 4 and 5 do not lie within the confines of any of the high spatial resolution radio continuum data that we have access to, and thus whether or not they have radio continuum emission is unknown. Five MYSOs (3, 9, 10, 11, and IRS9) are located within extended diffuse radio continuum emission, but there are no radio continuum peaks at their locations, and thus it is not clear if they are emitting radio continuum emission or not. For the eight MYSOs with no detectable radio emission (2, 6, and 8) or no peak above the extended continuum (3, 9, 10, 11, and IRS9) it is possible that they are in a very young state prior to the onset of a hypercompact \ion{H}{2} region \citep{2010ApJ...721..478H} and not observable via radio continuum emission.

\begin{figure*}[tb!]
\epsscale{1.15}
\plotone{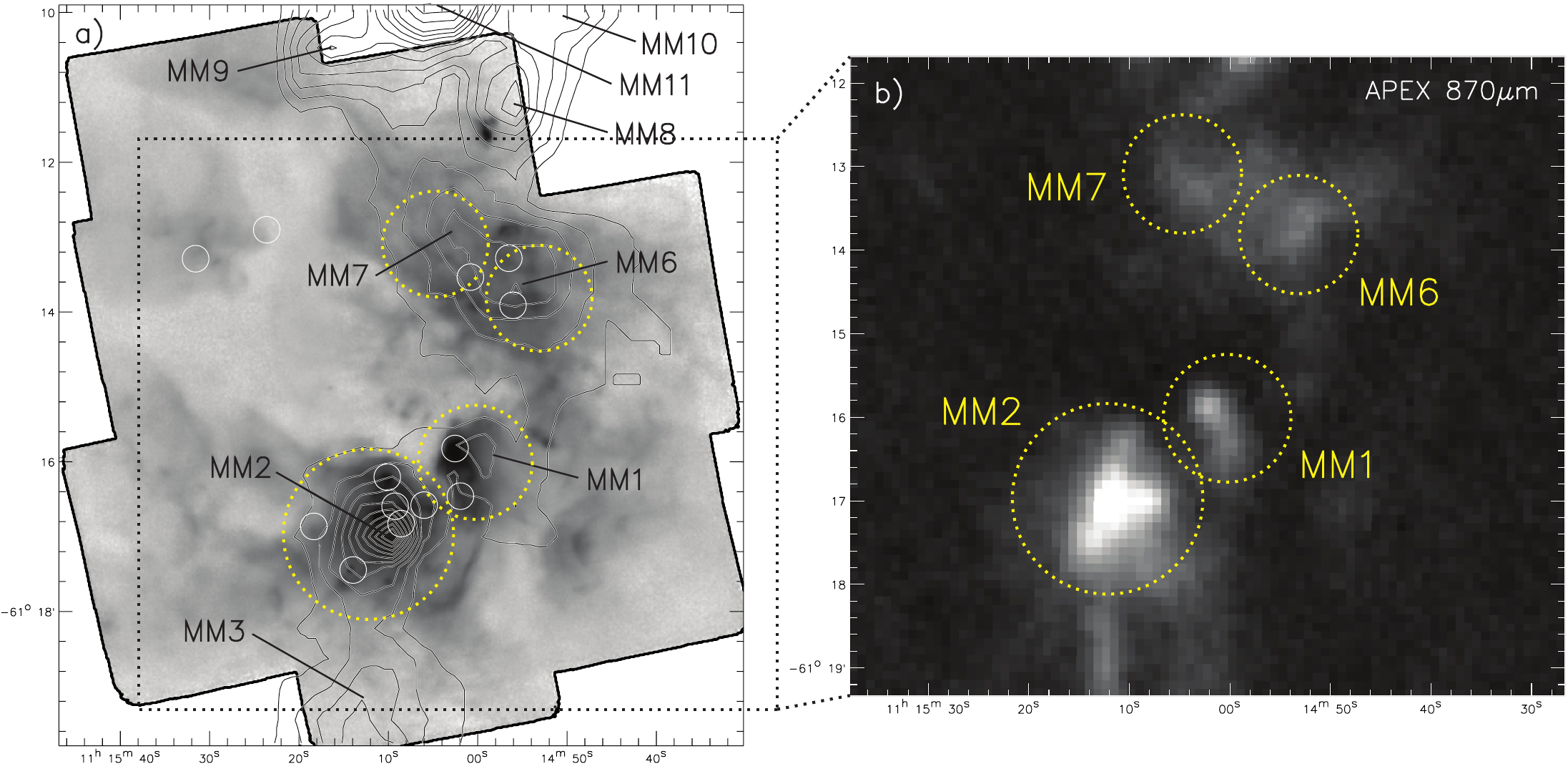}
\caption{Extended sub-regions of NGC~3603 used for evolutionary analyses. a) The background image is SOFIA 37\,$\mu$m and the black contours are the CS observations of \citet{2002A&A...394..253N} showing the major molecular clump locations (labeled MM1, MM2, etc.). White circles are the radio continuum peak locations from \citet{1999AJ....117.2902D}. Yellow dashed circles are the sub-regions defined in Table~\ref{tb:SOFIA_extended}. b) A slight zoom in on the dashed box in panel a) showing the field at 870\,$\mu$m as observed by APEX. Again, the size and locations of the sub-regions are marked. \label{fig:extend}}
\end{figure*}

Of all the MYSOs found in NGC~3603, the most massive is IRS9 with a best fit mass of 64\,$M_{\sun}$. Both IRS9 and source F top out at 64\,$M_{\sun}$ in the range of best fit model masses, meaning {\it no} source is fit with any model greater than this value. It is interesting that the most massive MYSO in the most powerful G\ion{H}{2} region in the Galaxy (in terms of Lyman continuum flux) is smaller than the most massive MYSOs seen in the next two most powerful G\ion{H}{2} regions we have studied, W49A (128\,$M_{\sun}$) and W51A:G49.5-0.4 (96\,$M_{\sun}$). Furthermore, compared to both W49A (24) and W51A:G49.5-0.4 (37), NGC~3603 has a dearth of MYSOs in general (18). Put another way, W49A is $2\times$ fainter and W51A:G49.5-0.4 is $4\times$ fainter in Lyman continuum flux than NGC~3603, but they contain $1.3\times$ and $2\times$ the number of MYSOs as NGC~3603, respectively. We will discuss a potential reason for this in Section \ref{sec:stats}. 

\subsection{Physical Properties of Extended Sub-Regions: Evolutionary Analysis Methodology and Kinematic Status}\label{sec:es}

\begin{deluxetable*}{lccrrrrrr}
\tabletypesize{\scriptsize}
\tablecolumns{9}
\tablewidth{0pt}
\tablecaption{SOFIA Observational Parameters of Sub-Regions Within NGC3603}
\tablehead{\colhead{  }&
            \colhead{  }&
            \colhead{  }&
           \multicolumn{3}{c}{${\rm 20\mu{m}}$}&
           \multicolumn{3}{c}{${\rm 37\mu{m}}$}\\
           \cmidrule(lr){4-6} \cmidrule(lr){7-9}\\
           \colhead{ Source }&
           \colhead{ R.A.}&
           \colhead{ Decl. }&           
           \colhead{ $R_{\rm int}$ } &
           \colhead{ $F_{\rm int}$ } &
           \colhead{ $F_{\rm int-bg}$ } &
           \colhead{ $R_{\rm int}$ } &
           \colhead{ $F_{\rm int}$ } &
           \colhead{ $F_{\rm int-bg}$ } \\
	   \colhead{  } &
           \colhead{(J2000) }&
           \colhead{(J2000) }& 	   
	   \colhead{ ($\arcsec$) } &
	   \colhead{ (Jy) } &
	  \colhead{ (Jy) } &
	   \colhead{ ($\arcsec$) } &
	   \colhead{ (Jy) } &
	   \colhead{ (Jy) } \\
}
\startdata
MM1	&	11:15:00.36	&	-61:16:00.6	&	46	&	4610	&	3640	&	46	&	6740	&	4640	\\
MM2	&	11:15:12.18	&	-61:16:59.0	&	69	&	8350	&	7000	&	69	&	15500	&	12500	\\
MM6	&	11:14:53.33	&	-61:13:49.3	&	43	&	1851	&	824	&	43	&	3610	&	1057	\\
MM7	&	11:15:05.07	&	-61:13:06.6	&	42	&	2096	&	746	&	42	&	4340	&	1334	\\
\enddata
\tablecomments{Sub-regions are defined as large, contiguous regions, as seen in the CS, CO, and 870\,$\mu$m  maps.}
\label{tb:SOFIA_extended}
\end{deluxetable*}

\begin{deluxetable*}{rccccccc}
\tabletypesize{\small}
\tablecolumns{8}
\tablewidth{0pt}
\tablecaption{Derived Parameters of Sub-Regions in NGC3603}\label{tb:esngc3603}
\tablehead{\colhead{   Source   }                                              &
           \colhead{  $M_{\rm vir}$   } &
           \colhead{  $M$   } &
           \colhead{ $L$ } &
           \colhead{  $T_{\rm cold}$  } &
           \colhead{  $T_{\rm warm}$  } &
           \colhead{  $L/M$  } &
           \colhead{ $\alpha_{\rm vir}$ }\\
	   \colhead{        } &
	   \colhead{ ($M_{\sun}$) } &
	   \colhead{ ($M_{\sun}$) } &
	   \colhead{ ($\times 10^4 L_{\sun}$) } &
	   \colhead{ (K) } &
	   \colhead{ (K) } &
       \colhead{  $L_{\sun}/M_{\sun}$  } &
       \colhead{     }
}
\startdata
    MM1 &  1369 &    212.9 &     242 &     149.9 & 224.7 &  5680 &   6.43\\
    MM2 &  3327 &     1560 &     625 &     142.5 & 230.5 &  2002 &   2.13\\
    MM6   & 455.8 &    264.9 &    80.2 &     128.2 & 214.3 &  1514 &   1.72\\
     MM7   &  1289 &    293.3 &     107 &     128.6 & 238.3 &  1821 &   4.39\\
\enddata
\end{deluxetable*}\label{tb:sede} 

The observations of CS by \citet{2002A&A...394..253N} and CO by \citet{2014ApJ...780...36F} show that there appears to be a filament of molecular material present here extended $\sim$7$\arcmin$ to the north and $\sim$6$\arcmin$ to the south of the HD~97950 cluster. In the CS observations of \citet{2002A&A...394..253N} they identify 13 molecular clumps in this filament, which they label MM1 to MM13, and show that the HD~97950 cluster lies near the center of (and presumably formed from) the filament. The clumps MM1 and MM2 lie just to the south of the HD~97950 cluster and are the closest in projection. Slightly farther away and to the north of the stellar cluster are molecular clumps MM6 and MM7. Together, these four clumps are coincident with the brightest near- to mid-infrared emitting dust of NGC~3603 and, in particular, the brightest areas seen in our SOFIA maps (Figure~\ref{fig:extend}). Using high spatial resolution 870\,$\mu$m data from the APEX telescope\footnote{Based on observations under program ID 081.F-9325(A). APEX is a collaboration between the Max-Planck-Institut fuer Radioastronomie, the European Southern Observatory, and the Onsala Observatory.}, and cross referencing it with $^{13}$CO$(1-0)$ archival data from the Mopra 22~m telescope of the Australia Telescope National Facility \citep[Project I.D. - M161;][]{2018ApJ...866...19B}, we were able to confirm the locations and extents of these molecular clumps (i.e., the extended sub-regions) contained in our SOFIA fields, and these are shown in Figure~\ref{fig:extend}. We tabulate the relevant information describing sub-region locations, photometric apertures used, and measured SOFIA fluxes in Table~\ref{tb:SOFIA_extended} (with Spitzer and Herschel photometric data given in Appendix~\ref{sec:appendixflux}).

As we did in our previous papers, we can use the SOFIA data along with gas kinematics information to study the evolutionary state of these extended sub-regions (assuming they are star-forming molecular clumps) and try to see if they give us clues as to how the region came together or evolved to its present state. In previous papers of this project, we conducted a comparative analysis between two independent tracers of molecular clump evolution: virial parameter and $L/M$, focusing on individual extended sources within G\ion{H}{2} regions. The analysis revealed a distinct positive correlation between the virial parameter ($\alpha_{\rm vir}$) and $L/M$ for the sub-regions. Higher values of both $\alpha_{\rm vir}$ and $L/M$ were interpreted as indicative of relatively older sub-regions.

We extended this analysis to the extended sub-regions within NGC\,3603. Overall our approach was similar to that used in \citetalias{2019ApJ...873...51L}, but we we highlight here the key steps. First, to derive the $L/M$ values we determined the masses using a pixel-by-pixel graybody fitting method, following the technique developed by \citet{2016ApJ...829L..19L}. To achieve higher angular resolution, we utilized the Herschel 160 to 500~$\mu$m images and convolved them to a common beam size of 36$\arcsec$ to obtain a `template temperature' ($T$) map. Subsequently, we applied this temperature map to the APEX 870~$\micron$ data for improved density mapping (i.e., $\sim$ 18$\arcsec$ resolution). Second, we calculated the bolometric luminosities ($L$) through a two-temperature graybody fit using the integrated total fluxes for each source in each Spitzer, SOFIA, and Herschel filter. Background flux estimation was performed using the data in an annulus outside of the photometric aperture of each extended source. 

Then, to derive the virial parameters ($\alpha_{\rm vir}$) of each extended sub-region, we followed the methods laid out in \citetalias{2019ApJ...873...51L}. In short, the virial analysis entails comparing the gravitational potential energy of extended sources to the total kinematic energy using the virial parameter ($\alpha_{\rm vir} = M_{\rm vir}/M$). In this study, the analysis involves calculating the latter from the full width at half maximum (FWHM) of the integrated $^{13}$CO$(1-0)$ line profiles of Mopra data for each of the clumps (see Equation 2 of \citetalias{2019ApJ...873...51L}). 

Table~\ref{tb:sede} summarizes the physical parameters we derived for each star-forming clump, including the virial mass ($M_{\rm vir}$), clump mass ($M$), bolometric luminosity ($L$), the derived warm and cold temperature components ($T_{\rm cold}$ and $T_{\rm warm}$), the luminosity-to-mass ratio ($L/M$), and the virial parameter ($\alpha_{\rm vir}$).

As shown in Table~\ref{tb:sede},  the extended sources within NGC\,3603 exhibit mass ranges from $212.9M_{\sun}$ to $1560M_{\sun}$, with an average mass ($\overline{M}$) of approximately $583M_{\sun}$. In comparison, W51A (\citetalias{2019ApJ...873...51L}), M17 (\citetalias{2020ApJ...888...98L}), and W49A (De Buizer et al. 2021; hereafter ``\citetalias{2021DeBuizer}'') have mean masses of approximately $3500M_{\sun}$, $2100M_{\sun}$, and $1400M_{\sun}$, respectively. It is important to highlight that despite being the most luminous G\ion{H}{2} region in the Milky Way in terms of Lyman continuum luminosity (\citealt{2004MNRAS.355..899C}; \citetalias{2022DeBuizer}), NGC\,3603 exhibits much smaller $\overline{M}$ values for its molecular clumps compared to other G\ion{H}{2} regions with similarly large $N_{LyC}$ values (i.e., W49A and W51A). 

\begin{figure*}[tb!]
\epsscale{0.65}
\plotone{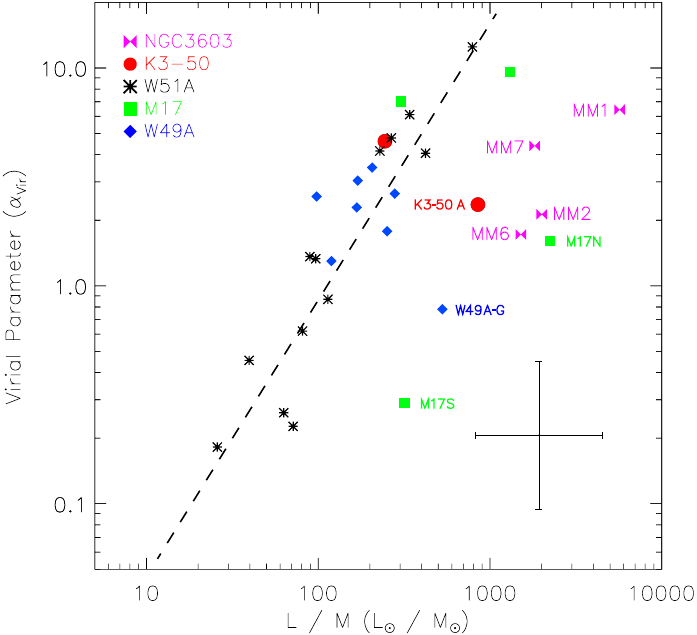}
\caption{\footnotesize Virial parameter ($\alpha_{\rm vir}$) vs. $L/M$ of all infrared sub-regions in all G\ion{H}{2} regions studied so far. Black asterisks are values for the sub-regions in W51A (i.e., both the G49.5-0.4 and G49.4-0.3 G\ion{H}{2} regions), and the  dashed line indicates the best line fit to the W51A data ($\alpha\sim$1.28 in log-space). Green squares show the sub-regions of M17, blue diamonds show the data for the sub-regions in W49A, and the red dots are for K3-50. All labeled sub-regions, including all sub-regions in NGC~3603, appear to have significantly higher $L/M$ values that place them to the right of the main trend. These regions are believed to have higher $L/M$ values due to contamination by external heating/ionization. The error bar at the bottom left shows the typical uncertainty (a factor of $\sim$2) in both $L/M$ and $\alpha_{\rm vir}$.} \label{fig:almplot}
\end{figure*}

This trend of lower mass and higher luminosity can be seen in our measured $L/M$ values. The minimum and maximum $L/M$ values for the extended sources in NGC\,3603 are approximately $1500\,L_{\sun}/M_{\sun}$ and $5700\,L_{\sun}/M_{\sun}$, respectively. In contrast, W51A, W49A, and M17 show $L/M$ ranges of $26 \lesssim{L/M} \lesssim 800\,L_{\sun}/M_{\sun}$, $50 \lesssim {L/M} \lesssim 500\,L_{\sun}/M_{\sun}$, and $300 \lesssim {L/M} \lesssim 2000\,L_{\sun}/M_{\sun}$, respectively. Since larger $L/M$ values denote more evolved sources \citep[e.g.,][]{2007ApJ...654..304K}, this suggests that, globally, NGC\,3603 is in a more advanced evolutionary stage compared to the other G\ion{H}{2} regions studied in this project (though as we will discuss below, there may be an additional reason for the higher $L/M$ values).

The derived dust temperatures based upon the graybody model employed also seem to indicate that NGC~3603 is an overall more evolved G\ion{H}{2} region than the others. In Table~\ref{tb:sede}, we list the $T_{\rm cold}$ and $T_{\rm warm}$ of the extended sources of NGC\,3603, which span along the ranges of $128~{\rm K}\lesssim T_{\rm cold}\lesssim150~{\rm K}$ and $214~{\rm K}\lesssim T_{\rm warm}\lesssim238~{\rm K}$, respectively. Compared to the other G\ion{H}{2} regions we have studied, the $T_{\rm cold}$ values for the NGC\,3603 extended sources are $50-70~{\rm K}$ higher, and higher temperatures are generally expected for more evolved clumps. This is also seen in \citetalias{2019ApJ...873...51L}, where the older population of molecular clumps in the G49.5–0.4 region (i.e., sources f to j) of W\,51\,A possess $\sim20-30~{\rm K}$ higher $T_{\rm cold}$ values compared to the younger clumps in the G49.5–0.3 and G49.5–0.4 regions. 

Our virial analysis of the extended sources was used to assess the degree of kinematic stability within G\ion{H}{2} regions and examine overall trends. The $\alpha_{\rm vir}$ values for NGC\,3603 ranged from 1.72 to 6.43. Among these, only one sub-region (MM6) is gravitationally bound ($\alpha_{\rm vir}<2$), while the remaining three sources are unbound ($\alpha_{\rm vir}>2$). Unlike other G\ion{H}{2} regions in this project, NGC\,3603 did not exhibit any self-collapsing molecular clumps ($\alpha_{\rm vir}<1$). Therefore, the overall $\alpha_{\rm vir}$ analysis aligns with the other analyses described above, indicating that NGC\,3603 contains a relatively older clump population compared to other G\ion{H}{2} regions.

To further investigate the evolutionary states of star-forming molecular clumps in NGC\,3603, we created a plot of $\alpha_{\rm vir}$ versus $L/M$, as depicted in Figure~\ref{fig:almplot}. We included clumps from previous studies in this project for comparison. Similar to W49A, the plot clearly illustrates that the range of relative ages of clumps in NGC\,3603 is smaller than in W51A and M17. It is also discernible that the extended sources of NGC\,3603 show higher $L/M$ values than those of W51A and W49A relative to their $\alpha_{\rm vir}$ value so that NGC\,3603 sources appear shifted to the right from the major trend of W51A and W49A in Figure~\ref{fig:almplot}. The $L/M$ analysis is supposed to utilize measurements of the intrinsic luminosity and mass of the molecular clump, however we argued in previous papers that the sources lying to the right of the main trend are likely to have heightened luminosity measurements (and therefore $L/M$ values) due to external heating. For instance, for the two extended sources of M\,17 (M17\,N and M17\,S) we suggested that external heating from the nearby evolved massive stellar cluster (NGC\,6618) was affecting the SED of the surrounding molecular clumps (\citetalias{2020ApJ...888...98L}). Similarly, the extended sources in NGC~3603 are all likely to be heavily influenced by its central stellar cluster, HD~97590, which in turn augments the $L/M$ values we derive. Therefore, in NGC~3603 the virial parameter may be a better indication of relative ages than the $L/M$ values. That being said, the location of all the data points in the upper right corner of Figure~\ref{fig:almplot} is consistent with the clumps in NGC~3603 all being relatively older. Furthermore, the values for $\alpha_{\rm vir}$ and $L/M$ for clumps MM2, MM6, and MM7 are practically the same given the errors (see the error bars in Figure~\ref{fig:almplot}), indicating the age spread of the clumps is also relatively small. 

\subsubsection{The Evolutionary History of NGC~3603}\label{sec:alm}

Our results from the compact source analyses show that there is a modest amount of massive star formation presently occurring, especially for a G\ion{H}{2} region the size and power of NGC~3603 (more on this in \S4.2.2). The star formation is also isolated to a small number of molecular cores (which are producing the MYSOs we see). However, none of the large molecular clumps that make up NGC~3603 appear to be globally collapsing, and therefore may not produce a substantial subsequent star formation population in the future.

In terms of the evolutionary history of NGC~3603, \citet{1989A&A...213...89M} was the first to claim that there may be signs of sequential triggered star formation within the G\ion{H}{2} region. They claimed that there is evidence of an older population of stars, pointing to the presence of several evolved stars in the area, like the blue supergiants Sher 18 and Sher 23 \citep{1997ApJ...475L..45B}, as well as Sher 25, which \citet{1989A&A...213...89M} estimated to have formed $\sim$10~Myr ago. \citet{1977ApJ...213..723F} also claimed that IRS4, 5, 6, and 15 are late-type K and M supergiant stars, which \citet{1989A&A...213...89M} estimated to be on the order of 7~Myr old. Separate from this older collection of stars, the HD~97950 cluster itself has been determined by many groups to be much younger, at around $\sim$1-3~Myr old \citep[e.g.,][]{2008AJ....135..878M, 2006AJ....132..253S, 2004AJ....127.1014S, 1999A&A...352L..69B, 1998ApJ...498..278E}. This led \citet{1989A&A...213...89M} to suggest that star formation occurred in two epochs (one $\sim$10~Myr ago and the other $\sim$1~Myr ago). Furthermore, \citet{1989A&A...213...89M} suggest that the older stars appear to be concentrated to the north, and so perhaps star formation may have proceeded from the north (i.e., around the MM1 and MM2 sub-regions) to south of the HD~97950 cluster (i.e., around the MM2 sub-region), and that this might indicate some form of sequential star formation.

\begin{deluxetable*}{lcccccc}
\tabletypesize{\small}
\tablecolumns{12}
\tablewidth{0pt}
\tablecaption{Infrared Observational Indicators of All Surveyed G\ion{H}{2} Regions to Date }\label{tb:class}
\tablehead{\colhead{Region} &
           \colhead{ log$N_{Lyc}$ } &
           \colhead{ G\ion{H}{2} } &
           \colhead{ No. Compact   } &
           \colhead{ No. Sub-  } &
           \colhead{ \%Flux in } &
           \colhead{  Highest Mass } \\
           \colhead{} &
           \colhead{ (log s$^{-1}$) } &
           \colhead{ Type } &
           \colhead{ Sources  } &
           \colhead{ Regions  } &
           \colhead{ Peak} &
           \colhead{ YSO ($M_{\sun}$)} 
}
\startdata
NGC~3603        & 51.61 &      Cavity &   18 &        4&       3 &       64 \\
W49A            & 51.42 & Distributed &   24 &      15 &      25 &      128 \\
W51A: G49.5-0.4 & 51.03 & Distributed &   37 &      10 &      20 &       96 \\
M17             & 51.02 &      Cavity &   16 &       4 &       5 &       64 \\
\hline
W51A: G49.4-0.3 & 50.22 & Distributed &   10 &       5 &      15 &       64 \\
DR7             & 50.10 &      Cavity &    4 &       1 &      15 &       16 \\
K3-50           & 50.07 & Distributed &   10 &       5 &      59 &       48 
\enddata
\end{deluxetable*}

Our evolutionary analyses show that clump MM6 to the west is the youngest, and clump MM1 in the southwest is oldest, with MM7 to the north and MM2 to the south being at some relative age in between. There is therefore no obvious age trend north-south, which seems to contradict the sequential north-south star formation scenario of \citet{1989A&A...213...89M}. Nor does there seem to be a trend east-west or inside-out either, signifying that there may be no globally triggered sequential star formation of any kind occurring in NGC~3603. Indeed, as we have seen from the GAIA distances to the IRS sources (Table\,\ref{tb:IRS}), they are not physically associated with NGC\,3603. Furthermore, looking at the the GAIA distances for the blue supergiants Sher\,25 (5873$_{-102}^{+30}$\,pc) and Sher\,18 (4274$_{-94}^{+54}$\,pc) shows that they too are not part of NGC\,3603. Thus, the fact that the majority of these evolved sources lie (in projection) more to the north of HD~97950 \citep[as pointed out by][]{1989A&A...213...89M} and seem to have formed at a different epoch compared to the cluster is irrelevant to the evolutionary history of NGC\, 3603.  

The second hypothesis for the evolution of NGC~3603 comes from \citet{2014ApJ...780...36F}, who suggest that the present state of NGC~3603 was the result of a cloud-cloud collision. They find evidence in their CO data for molecular material at two discrete velocities, one at 13 km$\cdot$s$^{-1}$ and the other at 28 km$\cdot$s$^{-1}$. The molecular material detected at these velocities covers almost the entire filament as seen in the \citet{2002A&A...394..253N} CS maps (i.e., the region covering MM1-MM13). \citet{2014ApJ...780...36F} claim that the two clouds collided with each other $\sim$1~Myr ago instigating a burst of star formation that led to the creation of most of the stars in the area, including the HD~97950 star cluster. 

The fact that we find a relative small age spread in the clump ages of NGC~3603 would seem to support the idea that the majority of star formation may have occurred around the same time, which may be consistent with the cloud-cloud collision scenario. The mid-infrared MYSOs we identified in this work, the YSOs found in the studies by others \citep[e.g.,][]{1977ApJ...213..723F, 2003AA...400..223N, 2015ApJ...799..100D}, the massive proplyd candidates \citep[][]{2002ApJ...571..366M, 2000AJ....119..292B}, as well as the presence of maser emission from water, methanol, and OH species \citep[e.g.,][]{2010MNRAS.406.1487B, 2004MNRAS.351..279C, 1989AuJPh..42..331C}, all provide evidence or recent star formation activity, however at a very modest level, as mentioned above. Furthermore, \citet{2014ApJ...780...36F} suggest that the moderate collection of older stars seen in the area is also evidence that there was only modest star formation activity before the cloud-cloud collision as well (and we additionally now know most of them are foreground stars).

\subsubsection{Two Types of GH II Regions?}\label{sec:stats}

In \citetalias{2022DeBuizer}, we proposed the utilization of various secondary criteria to potentially distinguish whether a given region qualifies as a genuine G\ion{H}{2} region or not, especially in cases where the $N_{LyC}$ value as measured from the cm radio continuum flux exhibits a considerable error or is in close proximity to the qualifying threshold. The suggested supplementary indicators were: the number of compact infrared sources within the region, the number of sub-regions it comprises, the proportion of the total infrared flux emanating from the brightest source within the region, and the mass of the most massive MYSO. In \citetalias{2023ApJ...949...82D}, we discuss that there was evidence suggesting that the population of G\ion{H}{2} regions may exhibit two distinct morphological variations, one characterized by dispersed radio sub-regions (i.e., `distributed-type') and the other marked by contiguous cavity structures (i.e., `cavity-type'). The evolutionary pathways, and consequently the observed properties, of a G\ion{H}{2} region may vary significantly between the two morphological classifications, hinting at the possibility that the aforementioned indicators could be an oversimplification. 

Indeed, NGC~3603 has a cavity-like morphology similar in appearance to the previously studied M17 and DR7 G\ion{H}{2} regions in our survey. We add the values for the supplementary indicators discussed above to those for the other sources we have observed in this survey with log$N_{LyC}>50.0$ and list them in Table~\ref{tb:class}. These regions are listed in order by $N_{LyC}$ value, and a horizontal line separates those with similar values, i.e. those with $51.02 < N_{LyC} < 51.61$ and those with $50.07 < N_{LyC} < 50.22$. For the four sources with a higher range of $N_{LyC}$ it can be seen that the two with a cavity-type morphology have similar but lower values in all indicators, when compared to those with a distributed-type morphology, which have higher values in all indicators. Likewise, for the three regions with a lower range of $N_{LyC}$, the cavity-type GHII region is a outlier with lower indicators as well. We do caution that these are small number statistics, and we will continue to investigate this issue as we continue our survey in future papers. 

Nevertheless, as first discussed in \citetalias{2023ApJ...949...82D}, these two morphological classes of G\ion{H}{2} region might point to a fundamentally different formation history. NGC~3603, along with M17 and DR7, are predominantly characterized by cavity structures, where a previously formed stellar cluster appears to be responsible for the vast majority of the ionization and dust heating within the cavity. In contrast, the other regions (W51A:G49.5-0.4, W51A:G49.4-0.3, W49A, and K3-50) consist of dusty, ionized sub-regions within larger molecular clumps with recent and ongoing massive star formation, contributing significantly to the overall Lyman continuum flux. In these regions, previous star formation likely plays a smaller role.

Cavity-type sources may have experienced intense star formation in the past, resulting in radiation pressure that initially prevented the formation of subsequent stars. Over time, this pressure cleared surrounding material until overdensities in the cavity walls became dense enough to locally collapse into new stars. For NGC~3603, M17, and DR7 it can be seen that MYSOs and compact sources are concentrated near these cavity walls. These regions may have fewer MYSOs overall, possibly indicating that they are more evolved G\ion{H}{2} regions or are in a transitional state between major star formation events. Moreover, cavity-type regions like NGC~3603 exhibit a lack of extended sub-regions, with most of the infrared and radio emission originating from the ridges outlining the cavity walls being carved out of the larger molecular cloud. As a result, the suggested observation indicators for G\ion{H}{2} regions that were suggested originally in \citetalias{2022DeBuizer}, may not be reliable indicators of G\ion{H}{2} region status for regions with a cavity-type morphology. As a consequence, the assessment of \citetalias{2022DeBuizer} that DR7 may not be a genuine G\ion{H}{2} region may not be correct, and evaluation of more sources will be required to better understand the observational properties of both G\ion{H}{2} morphological classes.

\section{Summary}\label{sec:sum}

We obtained SOFIA-FORCAST 20 and 37\,$\mu$m maps toward NGC~3603, covering the brightest central infrared-emitting area ($\sim$$8\farcm5\times8\farcm5$) of the G\ion{H}{2} region at $\lesssim$3$\arcsec$ spatial resolution. These infrared observations represent the best spatial resolution images yet of NGC~3603 at wavelengths more than 25\,$\mu$m. In order to examine the morphological and physical characteristics of the compact sources and extended sub-regions within NGC~3603, we compared these SOFIA-FORCAST images with earlier multi-wavelength studies spanning the near-infrared to radio wavelengths from various ground- and space-based telescopes. We used a MYSO SED model fitting algorithm to deduce properties of the compact infrared sources, and applied evolutionary analyses to the extended sub-regions within NGC~3603 under the assumption that they trace star-forming molecular clumps. We itemize below our main conclusions from this work.

1) The SOFIA 20 and 37\,$\mu$m images show similar large-scale structures. However, there is a distinct heating pattern that can be seen between the two wavelengths, with the 20\,$\mu$m emission more strongly tracing hotter dust nearer to the HD~97590 stellar cluster. At these wavelengths, emission can also be seen coming from the various known pillars and trunks, all of which point back to the  central stellar cluster.

2) There are three proplyd-candidates previously identified in NGC~3603, however it was unclear if they possessed circumstellar disks like the archetypical Orion proplyds. They are also more than an order of magnitude larger and have radio continuum emission, though it was unclear if that emission is due to external or internal ionization. With SOFIA, all three sources are detected in the infrared, suggesting that they do indeed possess circumstellar dust of some kind, likely from both a disk and envelope. While the Orion proplyds contain low-mass stars, our SED modeling shows that these sources are indeed likely to be massive, and thus it is unclear if they are high-mass proplyds analogs or cometary-shaped UC\ion{H}{2} regions.

3) Sher 25 is an evolved blue supergiant in the foreground of NGC~3603 that is surrounded by a circumstellar ring with bipolar outflow clouds that are believed to be part of a larger hourglass-shaped circumstellar nebula. Our SOFIA 20\,$\mu$m image shows dust emission present around the entire circumstellar ring with infrared dust clumps well-correlated with the clumps seen in the Hubble H$\alpha$ image. The outflow clouds are more prominent at 37\,$\mu$m and we argue that their dust is cooler than the dust in the circumstellar ring. We also detect faint emission from the entire hourglass nebula structure at 20\,$\mu$m, confirming its presence.

4) \citet{1977ApJ...213..723F} identified and labeled infrared sources in the field of NGC~3603, with IRS1--3 being identified in their 10 and 20\,$\mu$m images, and IRS4--15 in their 2\,$\mu$m images. Our SOFIA 20 and 37\,$\mu$m images show no source at the location of IRS3, and no source can be seen in the Spitzer or Herschel data either. IRS4--15 are quoted with 10$\arcsec$ astrometric accuracy by \citet{1977ApJ...213..723F}, and we use 2MASS data to identify these sources and refine their coordinates. IRS4 and IRS9 are the only NIR-defined sources detected in our SOFIA mid-infrared data. IRS4 is the brightest stellar object in all of NGC~3603 in the NIR but decreases in flux quickly with increased wavelength. Its distance, as deduced from GAIA parallax measurements, is only 3.4\,kpc away, and it is therefore a foreground stellar source and not a YSO. Similarly, we find that most of the remaining IRS sources with no SOFIA detections but for which GAIA distances are available are also foreground stars unrelated physically to NGC~3603.

5) We found 18 compact sources in the SOFIA data, and 8 are newly identified here. Sixteen sources had sufficient photometry data from SOFIA, Spitzer-IRAC, and Herschel-PACS to construct SEDs. We fit those SEDs with young stellar object models and found 14 of the 16 sources (88\%) are likely to be massive young stellar objects, several of which are identified as such in this work for the first time. Sources 1 and 7, however, appear to be better fit with single temperature blackbodies, and we suggest that they maybe be externally heated dust shells or knots. Fourteen of the total 18 compact sources do not have radio continuum components, implying very young states of formation. However, two of the remaining four sources are located outside our radio continuum maps, so whether they have associated radio continuum emission is unknown.

6) We calculated the luminosity-to-mass ratio ($L/M$) and virial parameters ($\alpha_{\rm vir}$) of the extended sub-regions of NGC~3603 to estimate their relative ages. Unlike other G\ion{H}{2} regions in this project, NGC~3603 does not seem to have any self-collapsing molecular clumps ($\alpha_{\rm vir} < 1$), indicating that it is an older G\ion{H}{2} region overall. Consistent with this, NGC~3603 was also found to have a much higher range of $L/M$ ratios compared to the other G\ion{H}{2} regions (though external heating complicates the interpretation). Further evidence of this comes from our derivations of dust temperature, which show that the star-forming clumps in NGC~3603 have much higher cold dust temperatures, which are generally expected for more evolved molecular clumps. 

7) The absence of discernible age trends in various directions of the molecular clumps within NGC~3603 suggests that there is no globally triggered sequential star formation occurring, but instead, the relatively small age spread among clumps supports the idea of synchronized star formation, perhaps consistent with the cloud-cloud collision formation scenario.

8) Though it is the most powerful G\ion{H}{2} region in the Galaxy based upon its Lyman continuum photon rate, the most massive MYSO in NGC~3603 is estimated to be only 64\,$M_{\sun}$, which is significantly smaller than the most massive MYSOs seen in the next two most powerful G\ion{H}{2} regions we have studied, W49A and W51A:G49.5-0.4. It also has far fewer MYSOs and extended infrared sub-regions. We argue that there are two classes of G\ion{H}{2} region, and that NGC~3603 belongs to the `cavity-type', which tend to have fewer YSOs, sub-regions, and more modest MYSO masses, whereas W49A and W51A:G49.5-0.4 belong to a `distributed-type' with very different observational properties in the infrared. 

\acknowledgments
We would like to thank the anonymous referee for their helpful suggestions which improved the final version of this manuscript. This research is based on archival data from the NASA/DLR Stratospheric Observatory for Infrared Astronomy (\textit{SOFIA}). \textit{SOFIA} is jointly operated by the Universities Space Research Association, Inc. (USRA), under NASA contract NAS2-97001, and the Deutsches \textit{SOFIA} Institut (DSI) under DLR contract 50 OK 0901 to the University of Stuttgart. This work is also based in part on archival data obtained with the \textit{Spitzer Space Telescope}, which is operated by the Jet Propulsion Laboratory, California Institute of Technology under a contract with NASA. This work is also based in part on archival data obtained with \textit{Herschel}, an European Space Agency (ESA) space observatory with science instruments provided by European-led Principal Investigator consortia and with important participation from NASA. WL is supported by Caltech/IPAC under Contract No. 80GSFC21R0032 with the National Aeronautics and Space Administration.

\clearpage

\appendix
\section{Data release}

The FITS images used in this study are publicly available at: {\it https://dataverse.harvard.edu/dataverse/SOFIA-GHII}. 

The data include the \textit{SOFIA} FORCAST 20 and 37\,$\mu$m final image mosaics of NGC~3603 and their exposure maps.

\section{Distance to NGC~3603}\label{sec:appendixdist}
The distance to NGC~3603 of 7.2$\pm$0.1~kpc from \citet{2019MNRAS.486.1034D} was based upon parallax measurements from the GAIA mission's second data release (DR2). A more recent calculation based upon GAIA Early Data Release 3 (EDR3) data by \citet{2022A&A...657A.131M} yields a distance of $7130^{+590}_{-500}$ pc. Since these two distance measurements are equal to within their errors, we adopt here the value from \citet{2019MNRAS.486.1034D} because it is quoted with smaller uncertainty. However, an even more recent result based upon GAIA ER3 data published just prior this article by \citet{2023arXiv230806164W} yields a closer distance with equally small uncertainty of 6250$\pm$150~pc. To derive this value, \citet{2023arXiv230806164W} only used stars from an $r\sim1\arcmin$ area centered on HD~97950, and rejected many stars previously included in the distance analyses of others on the basis of potentially discrepant proper motions and/or the fact that the stars did not match the 1-2~Myr isochrone of Melena et al. (2008). In the end, only 10 stars were used to determine the distance to NGC~3603 (compared to 288 stars used by \citealt{2019MNRAS.486.1034D} and 166 by \citealt{2022A&A...657A.131M}). Interestingly, \citet{2019MNRAS.486.1034D} also derive a distance using the GAIA DR2 data for just the inner 1$\arcmin$ of NGC~3603 and derive a much larger value of 8.2$\pm$0.4~kpc using the parallaxes of just 30 stars. Moreover, Melena et al. (2008) summarize many of the distance measurements to NGC~3603, and a value of 6250~pc would be placed among the nearest distances reported, but consistent to within the errors of most measurements, which average closer to 7~kpc. 

We reran the SED fitting algorithm for all compact sources under the assumption that the distance was at 6250~pc (instead of 7200~pc) and found that our results did not change significantly. The infrared sources identified in Table~\ref{tb:sed} as MYSOs are all still MYSOs at this slightly closer distance. The main differences are that, at the closer distance, IRS9 has a best fit model of only $32~M_{\sun}$ (rather than $64~M_{\sun}$), and the range of fits for all sources tops out at $32~M_{\sun}$ (again rather than $64~M_{\sun}$). In general, the range of A$_V$ of the best fit SED models is lower at the closer distance, and either the lower limit or upper limit value in the mass range decreases by one mass step (the SED models have discrete masses of 2, 4, 8, 12, 16, 24, 32, 48, 64, 128~$M_{\sun}$). However, at this closer distance, still no source has a lower limit mass below $8~M_{\sun}$. The mass derived from the best model fit for half of the sources (i.e., 8 of 16) changes (either higher or lower) by one mass step, while the other half are unaffected.

The results of the extended source analyses are also mildly affected, with calculations at the nearer distance yielding $\alpha_{\rm vir}$ and $L/M$ values still within our quoted uncertainty level (a factor of two) for all extended sources. Specifically, at a distance of 6250~pc, the $\alpha_{\rm vir}$ values of each extended source increases by $\sim 16\%$ while $L/M$ remains the same (both $L$ and $M$ decrease by $\sim 26\%$ on each source). The virial parameters show the range of 2.00$\lesssim\alpha_{\rm vir}\lesssim$7.46 at 6250~pc while the range is 1.72$\lesssim\alpha_{\rm vir}\lesssim$6.43 at our adopted distance. The overall trend seen in Figure~\ref{fig:almplot} stays the same for $\alpha_{\rm vir}$ vs. $L/M$ at the 6250~pc distance, leaving the conclusions from the discussion in \S\ref{sec:alm} unaffected.

\section{Additional Photometry of Compact and Extended Sources in NGC~3603}\label{sec:appendixflux}

In addition to the fluxes derived from the SOFIA-FORCAST data, we used derived photometry data in our SED analyses for our sources from both Spitzer-IRAC and Herschel-PACS. 

\begin{deluxetable*}{lrrrrrrrrrrrr}
\tabletypesize{\scriptsize}
\tablecaption{Spitzer-IRAC Observational Parameters of Compact Sources in NGC 3603}
\tablehead{\colhead{  }&
           \multicolumn{3}{c}{${\rm 3.6\mu{m}}$}&
           \multicolumn{3}{c}{${\rm 4.5\mu{m}}$}&
           \multicolumn{3}{c}{${\rm 5.8\mu{m}}$}&
           \multicolumn{3}{c}{${\rm 8.0\mu{m}}$}\\
           \cmidrule(lr){2-4} \cmidrule(lr){5-7} \cmidrule(lr){8-10} \cmidrule(lr){11-13}\\
           \colhead{ Source }&
           \colhead{ $R_{\rm int}$ } &
           \colhead{ $F_{\rm int}$ } &
           \colhead{ $F_{\rm int-bg}$ } &
                      \colhead{ $R_{\rm int}$ } &
           \colhead{ $F_{\rm int}$ } &
           \colhead{ $F_{\rm int-bg}$ } &
                      \colhead{ $R_{\rm int}$ } &
           \colhead{ $F_{\rm int}$ } &
           \colhead{ $F_{\rm int-bg}$ } &
                      \colhead{ $R_{\rm int}$ } &
           \colhead{ $F_{\rm int}$ } &
           \colhead{ $F_{\rm int-bg}$ } \\
	   \colhead{  } &
	   \colhead{ ($\arcsec$) } &
	   \colhead{ (mJy) } &
	   \colhead{ (mJy) } &
	   \colhead{ ($\arcsec$) } &
	   \colhead{ (mJy) } &
	   \colhead{ (mJy) } &
	   \colhead{ ($\arcsec$) } &
	   \colhead{ (mJy) } &
	   \colhead{ (mJy) } &
	   \colhead{ ($\arcsec$) } &
	   \colhead{ (mJy) } &
	   \colhead{ (mJy) } \\
}
\startdata
1	&	6.0	    &	38.4	&	13.0	&	7.2 	&	109	    &	54.1	&	7.2	    &	394	    &	250	    &	8.4	    &	2820	&	1740	\\
2	&	16	&	630	&	373	&	16	&	750	&	324	&	16	&	3290	&	1860	&	16	&	9280	&	4990	\\
3	&	12  	&	661	    &	237	    &	12  	&	792	    &	249	    &	12	    &	4250	&	1820	&	12	    &	8570	&	2900	\\
4	&	4.2	    &	47.9	&	7.75	&	4.2	    &	51.0	&	12.8	&	4.2	    &	423	    &	104	    &	4.2	    &	1070	&	341	    \\
5	&	4.8	    &	245	    &	185	    &	4.8	    &	365	    &	300	    &	7.2	    &	2280	&	1480	&	7.2	    &	5920	&	4090	\\
6	&	5.4	    &	88.8	&	45.7	&	5.4	    &	96.7	&	40.1	&	5.4	    &	446	    &	235	    &	5.4	    &	1070	&	499	    \\
7	&	3.0	    &	9.46	&	0.703	&	3.6	    &	22.8	&	5.72	&	3.6	    &	76.7	&	16.9	&	4.2	    &	396	    &	139	    \\
8	&	4.2	    &	61.4	&	21.9	&	4.2	    &	80.5	&	40.5	&	4.2	    &	290	    &	76.0	&	4.2	    &	969	    &	234	    \\
9	&	5.4	    &	149	    &	33.0	&	5.4	    &	160	    &	35.3	&	6.0	    &	1310	&	227	    &	6.0	    &	4420	&	748	    \\
10	&	3.6	    &	67.7	&	15.5	&	4.8	    &	189	    &	64.6	&	4.8	    &	642	    &	114	    &	4.2	    &	1730	&	71.7	\\
11	&	7.2	    &	321	    &	181	    &	7.2	    &	343	    &	184	    &	9.0	    &	2750	&	1350	&	9.0	    &	8660	&	3610	\\
12	&	\nodata	&	sat	    &	sat	    &	\nodata	&	sat	    &	sat	    &	\nodata	&	sat	    &	sat	    &	\nodata	&	sat	    &	sat	    \\
IRS4	&	\nodata	&	sat	    &	sat	    &	\nodata	&	sat	    &	sat	    &	\nodata	&	sat	    &	sat	    &	\nodata	&	sat	    &	sat	    \\
IRS9	&	3.0	    &	615	    &	540	    &	3.0	    &	814	    &	706	    &	3.0	    &	1780	&	1250	&	\nodata	&	sat	    &	sat	    \\
F	&	7.8	    &	212	    &	74.1	&	7.8	    &	515	    &	202	    &	7.8	    &	1760	&	522	    &	7.8	    &	11000	&	2470	\\
P1	&	3.6	    &	26.0    &	13.0	&	3.6	    &	38.0	&	16.6	&	4.2	    &	146	    &	42.4	&	4.8	    &	900	    &	222	    \\
P1	&	3.6	&	26	&	13	&	3.6	&	38	&	16.6	&	4.2	&	146	&	42.4	&	4.8	&	900	&	222	\\
P2	&	9.0	&	170	&	73.3	&	9.0	&	223	&	81.9	&	9.0	&	803	&	275	&	9.0	&	4570	&	2170	\\
P3	&	4.8	&	40.5	&	12.7	&	4.8	&	60	&	17.6	&	4.8	&	157	&	35.3	&	\nodata	&	sat	&	sat	
\enddata
\tablecomments{Entries with ``sat'' means the sources are themselves saturated in that band or are affected by array saturation effects from nearby bright sources. If they are saturated themselves, we use the point source saturation fluxes of 190, 200, 1400, and 740\,mJy at 3.6, 4.5, 5.8, and 8.0\,$\mu$m, respectively (from the Spitzer Observer’s Manual, Version 7.1.), as lower limits in the SED modeling.}
\label{tb:IRAC}
\end{deluxetable*}

\begin{deluxetable*}{lrrrrrrrrrrrr}
\tabletypesize{\scriptsize}
\tablecaption{Spitzer-IRAC Observational Parameters of Sub-Regions in NGC 3603}
\tablehead{\colhead{  }&
           \multicolumn{3}{c}{${\rm 3.6\mu{m}}$}&
           \multicolumn{3}{c}{${\rm 4.5\mu{m}}$}&
           \multicolumn{3}{c}{${\rm 5.8\mu{m}}$}&
           \multicolumn{3}{c}{${\rm 8.0\mu{m}}$}\\
           \cmidrule(lr){2-4} \cmidrule(lr){5-7} \cmidrule(lr){8-10} \cmidrule(lr){11-13}\\
           \colhead{ Source }&
           \colhead{ $R_{\rm int}$ } &
           \colhead{ $F_{\rm int}$ } &
           \colhead{ $F_{\rm int-bg}$ } &
                      \colhead{ $R_{\rm int}$ } &
           \colhead{ $F_{\rm int}$ } &
           \colhead{ $F_{\rm int-bg}$ } &
                      \colhead{ $R_{\rm int}$ } &
           \colhead{ $F_{\rm int}$ } &
           \colhead{ $F_{\rm int-bg}$ } &
                      \colhead{ $R_{\rm int}$ } &
           \colhead{ $F_{\rm int}$ } &
           \colhead{ $F_{\rm int-bg}$ } \\
	   \colhead{  } &
	   \colhead{ ($\arcsec$) } &
	   \colhead{ (Jy) } &
	   \colhead{ (Jy) } &
	   \colhead{ ($\arcsec$) } &
	   \colhead{ (Jy) } &
	   \colhead{ (Jy) } &
	   \colhead{ ($\arcsec$) } &
	   \colhead{ (Jy) } &
	   \colhead{ (Jy) } &
	   \colhead{ ($\arcsec$) } &
	   \colhead{ (Jy) } &
	   \colhead{ (Jy) } \\
}
\startdata
MM1	&	46.2	&	5.78	&	3.27	&	46.2	&	10.4	&	7.32	&	46.2	&	39.2	&	28.2	&	\nodata	&	sat	&	sat	\\
MM2	&	69.0	&	17.6	&	12.5	&	69.0	&	25.4	&	18.9	&	69.0	&	115	    &	90.7	&	\nodata	&	sat	&	sat	\\
MM6	&	43.0	&	3.61	&	1.33	&	43.0	&	5.07	&	1.17	&	43.0	&	18.9	&	2.88	&	\nodata	&	sat	&	sat	\\
MM7	&	42.2	&	6.73	&	4.14	&	42.2	&	7.73	&	4.70	&	42.2	&	31.3	&	9.95	&	\nodata	&	sat	&	sat	\\
\enddata
\tablecomments{Entries with ``sat'' means the sub-regions are affected by array saturation effects from bright sources. }
\label{tb:IRAC_extended}
\end{deluxetable*}

\begin{deluxetable*}{lrrrrrr}
\tabletypesize{\scriptsize}
\tablecolumns{7}
\tablewidth{0pt}
\tablecaption{Herschel-PACS Observational Parameters of Compact Sources in NGC3603}
\tablehead{\colhead{  }&
           \multicolumn{3}{c}{${\rm 70\mu{m}}$}&
           \multicolumn{3}{c}{${\rm 160\mu{m}}$}\\
           \cmidrule(lr){2-4} \cmidrule(lr){5-7}\\
           \colhead{ Source }&
           \colhead{ $R_{\rm int}$ } &
           \colhead{ $F_{\rm int}$ } &
           \colhead{ $F_{\rm int-bg}$ } &
           \colhead{ $R_{\rm int}$ } &
           \colhead{ $F_{\rm int}$ } &
           \colhead{ $F_{\rm int-bg}$ } \\
	   \colhead{  } &
	   \colhead{ ($\arcsec$) } &
	   \colhead{ (Jy) } &
	  \colhead{ (Jy) } &
	   \colhead{ ($\arcsec$) } &
	   \colhead{ (Jy) } &
	   \colhead{ (Jy) } \\
}
\startdata
1	&	16.0	&	191	&	\nodata	&	22.5	&	133	&	\nodata	\\
2	&	16.0	&	1070	&	\nodata	&	22.5	&	598	&	\nodata	\\
3	&	16.0	&	889	&	291	&	22.5	&	267	&	97.1	\\
4	&	16.0	&	830	&	\nodata	&	22.5	&	926	&	\nodata	\\
5	&	22.5	&	1560	&	939	&	22.5	&	894	&	\nodata	\\
6	&	16.0	&	496	&	\nodata	&	22.5	&	373	&	\nodata	\\
7	&	16.0	&	157	&	\nodata	&	22.5	&	114	&	\nodata	\\
8	&	16.0	&	392	&	\nodata	&	22.5	&	254	&	\nodata	\\
9	&	16.0	&	1230	&	\nodata	&	22.5	&	1120	&	\nodata	\\
10	&	16.0	&	1450	&	\nodata	&	22.5	&	1340	&	\nodata	\\
11	&	16.0	&	1190	&	\nodata	&	22.5	&	1140	&	\nodata	\\
12	&	16.0	&	83.2	&	\nodata	&	22.5	&	69.0	&	\nodata	\\
IRS4	&	16.0	&	389	&	\nodata	&	22.5	&	241	&	\nodata	\\
IRS9	&	16.0	&	2240	&	\nodata	&	22.5	&	1590	&	\nodata	\\
F	&	16.0	&	569	&	\nodata	&	22.5	&	331	&	\nodata	\\
P1	&	16.0	&	249	&	\nodata	&	22.5	&	178	&	\nodata	\\
P2	&	16.0	&	358	&	105	&	22.5	&	288	&	\nodata	\\
P3	&	16.0	&	250	&	\nodata	&	22.5	&	165	&	\nodata	\\
\enddata
\tablecomments{If there is no $F_{\rm int-bg}$ value for a source, then the source is not well resolved from other nearby sources and/or extended emission. For these sources, the $F_{\rm int}$ value is used as the upper limit in the SED modeling.}
\label{tb:PACS}
\end{deluxetable*}

\begin{deluxetable*}{lrrrrrr}
\tabletypesize{\scriptsize}
\tablecolumns{7}
\tablewidth{0pt}
\tablecaption{Herschel-PACS Observational Parameters of Sub-Regions in NGC3603}
\tablehead{\colhead{  }&
           \multicolumn{3}{c}{${\rm 70\mu{m}}$}&
           \multicolumn{3}{c}{${\rm 160\mu{m}}$}\\
           \cmidrule(lr){2-4} \cmidrule(lr){5-7}\\
           \colhead{ Source }&
           \colhead{ $R_{\rm int}$ } &
           \colhead{ $F_{\rm int}$ } &
           \colhead{ $F_{\rm int-bg}$ } &
           \colhead{ $R_{\rm int}$ } &
           \colhead{ $F_{\rm int}$ } &
           \colhead{ $F_{\rm int-bg}$ } \\
	   \colhead{  } &
	   \colhead{ ($\arcsec$) } &
	   \colhead{ (Jy) } &
	  \colhead{ (Jy) } &
	   \colhead{ ($\arcsec$) } &
	   \colhead{ (Jy) } &
	   \colhead{ (Jy) } \\
}
\startdata
MM1	&	44.8	&	5010	&	2840	&	44.8	&	1980	&	891	\\
MM2	&	70.4	&	17600	&	13600	&	70.4	&	7850	&	4710	\\
MM6	&	43.0	&	3900	&	1600	&	43.0	&	1470	&	639	\\
MM7	&	42.4	&	4950	&	2790	&	42.2	&	1660	&	848	\\
\enddata
\label{tb:PACS_extended}
\end{deluxetable*}

\begin{deluxetable*}{lrrrrrrrrr}
\tabletypesize{\scriptsize}
\tablecolumns{10}
\tablewidth{0pt}
\tablecaption{Herschel-SPIRE Observational Parameters of Sub-Regions in NGC3603}
\tablehead{\colhead{  }&
           \multicolumn{3}{c}{${\rm 250\mu{m}}$}&
           \multicolumn{3}{c}{${\rm 350\mu{m}}$}&
           \multicolumn{3}{c}{${\rm 500\mu{m}}$}\\
           \cmidrule(lr){2-4} \cmidrule(lr){5-7} \cmidrule(lr){8-10}\\
           \colhead{ Source }&
           \colhead{ $R_{\rm int}$ } &
           \colhead{ $F_{\rm int}$ } &
           \colhead{ $F_{\rm int-bg}$ } &
           \colhead{ $R_{\rm int}$ } &
           \colhead{ $F_{\rm int}$ } &
           \colhead{ $F_{\rm int-bg}$ } &
           \colhead{ $R_{\rm int}$ } &
           \colhead{ $F_{\rm int}$ } &
           \colhead{ $F_{\rm int-bg}$ } \\
	   \colhead{  } &
          \colhead{ ($\arcsec$) } &
	   \colhead{ (Jy) } &
	   \colhead{ (Jy) } &
	   \colhead{ ($\arcsec$) } &
	   \colhead{ (Jy) } &
	   \colhead{ (Jy) } &
	   \colhead{ ($\arcsec$) } &
	   \colhead{ (Jy) } &
	   \colhead{ (Jy) } \\
}
\startdata
MM1	&	46.2	&	551	&	216	&	44.8	&	182	&	49.6	&	42.2	&	72.6	&	27.7 \\
MM2	&	69.0	&	2670	&	2120	&	70.4	&	958	&	738 &	42.2	&	343	&	216 \\
MM6	&	43.0	&	504	&	232	&	43.0	&	188	&	81.8	&	42.2	&	80.9	&	14.8 \\
MM7	&	42.4	&	583	&	161	&	42.2	&	208	&	106	&	42.2	&	77.2	&	37.2 \\
\enddata
\label{tb:PACS_extended}
\end{deluxetable*}


\clearpage

\begin{thebibliography}{}

\bibitem[Bally et al.(2000)]{2000AJ....119.2919B} Bally, J., O'Dell, C.~R., \& McCaughrean, M.~J.\ 2000, \aj, 119, 2919. doi:10.1086/301385
\bibitem[Barnes et al.(2018)]{2018ApJ...866...19B} Barnes, P.~J., Hernandez, A.~K., Muller, E., et al.\ 2018, \apj, 866, 19. doi:10.3847/1538-4357/aad4ab
\bibitem[Bertoldi \& McKee(1992)]{1992ApJ...395..140B} Bertoldi, F. \& McKee, C.~F.\ 1992, \apj, 395, 140
\bibitem[Brandl et al.(1999)]{1999A&A...352L..69B} Brandl, B., Brandner, W., Eisenhauer, F., et al.\ 1999, \aap, 352, L69
\bibitem[Brandner et al.(1997a)]{1997ApJ...475L..45B} Brandner, W., Grebel, E.~K., Chu, Y.-H., et al.\ 1997a, \apjl, 475, L45. doi:10.1086/310460
\bibitem[Brandner et al.(1997b)]{1997ApJ...489L.153B} Brandner, W., Chu, Y.-H., Eisenhauer, F., et al.\ 1997b, \apjl, 489, L153. doi:10.1086/316795
\bibitem[Brandner et al.(2000)]{2000AJ....119..292B} Brandner, W., Grebel, E.~K., Chu, Y.-H., et al.\ 2000, \aj, 119, 292. doi:10.1086/301192
\bibitem[Breen et al.(2010)]{2010MNRAS.406.1487B} Breen, S.~L., Caswell, J.~L., Ellingsen, S.~P., et al.\ 2010, \mnras, 406, 1487. doi:10.1111/j.1365-2966.2010.16791.x
\bibitem[Burrows et al.(1995)]{1995ApJ...452..680B} Burrows, C.~J., Krist, J., Hester, J.~J., et al.\ 1995, \apj, 452, 680. doi:10.1086/176339
\bibitem[Caswell et al.(1989)]{1989AuJPh..42..331C} Caswell, J.~L., Batchelor, R.~A., Forster, J.~R., et al.\ 1989, Australian Journal of Physics, 42, 331. doi:10.1071/PH890331
\bibitem[Caswell(2004)]{2004MNRAS.351..279C} Caswell, J.~L.\ 2004, \mnras, 351, 279. doi:10.1111/j.1365-2966.2004.07784.x
\bibitem[Conti \& Crowther(2004)]{2004MNRAS.355..899C} Conti, P.~S. \& Crowther, P.~A.\ 2004, \mnras, 355, 899. doi:10.1111/j.1365-2966.2004.08367.x
\bibitem[Cutri et al.(2003)]{2003yCat.2246....0C} Cutri, R.~M., Skrutskie, M.~F., van Dyk, S., et al.\ 2003, VizieR Online Data Catalog, II/246
\bibitem[De Buizer et al.(2017)]{2017ApJ...843...33D} De Buizer, J.~M., Liu, M., Tan, J.~C., et al.\ 2017, \apj, 843, 33. doi:10.3847/1538-4357/aa74c8
\bibitem[De Buizer et al.(2021)]{2021DeBuizer} De Buizer, J.~M., Lim, W., Liu, M., et al.\ 2021, \apj, 923, 198. doi:10.3847/1538-4357/ac2d25 [Paper III]
\bibitem[De Buizer et al.(2022)]{2022DeBuizer} De Buizer, J.~M., Lim, W., Karnath, N., et al.\ 2022, \apj, 933, 60. doi:10.3847/1538-4357/ac6fd8 [Paper IV]
\bibitem[De Buizer et al.(2023)]{2023ApJ...949...82D} De Buizer, J.~M., Lim, W., Radomski, J.~T., et al.\ 2023, \apj, 949, 82. doi:10.3847/1538-4357/acc9c6 [Paper V]
\bibitem[De Pree et al.(1999)]{1999AJ....117.2902D} De Pree, C.~G., Nysewander, M.~C., \& Goss, W.~M.\ 1999, \aj, 117, 2902. doi:10.1086/300892
\bibitem[Di Cecco et al.(2015)]{2015ApJ...799..100D} Di Cecco, A., Faustini, F., Paresce, F., et al.\ 2015, \apj, 799, 100. doi:10.1088/0004-637X/799/1/100
\bibitem[Drew et al.(2019)]{2019MNRAS.486.1034D} Drew, J.~E., Mongui{\'o}, M., \& Wright, N.~J.\ 2019, \mnras, 486, 1034. doi:10.1093/mnras/stz864
\bibitem[Eisenhauer et al.(1998)]{1998ApJ...498..278E} Eisenhauer, F., Quirrenbach, A., Zinnecker, H., et al.\ 1998, \apj, 498, 278. doi:10.1086/305552
\bibitem[Frogel et al.(1977)]{1977ApJ...213..723F} Frogel, J.~A., Persson, S.~E., \& Aaronson, M.\ 1977, \apj, 213, 723. doi:10.1086/155203
\bibitem[Fukui et al.(2014)]{2014ApJ...780...36F} Fukui, Y., Ohama, A., Hanaoka, N., et al.\ 2014, \apj, 780, 36. doi:10.1088/0004-637X/780/1/36
\bibitem[Gaia Collaboration(2020)]{2020yCat.1350....0G} Gaia Collaboration\ 2020, VizieR Online Data Catalog, I/350
\bibitem[Goss \& Radhakrishnan(1969)]{1969ApL.....4..199G} Goss, W.~M. \& Radhakrishnan, V.\ 1969, \aplett, 4, 199
\bibitem[Gutermuth et al.(2009)]{2009ApJS..184...18G} Gutermuth, R. A., Megeath, S. T., Myers, P. C. et al., 2009, ApJS, 184, 18
\bibitem[Gvaramadze et al.(2013)]{2013MNRAS.430L..20G} Gvaramadze, V.~V., Kniazev, A.~Y., Chene, A.-N., et al.\ 2013, \mnras, 430, L20. doi:10.1093/mnrasl/sls041
\bibitem[Hendry et al.(2008)]{2008MNRAS.388.1127H} Hendry, M.~A., Smartt, S.~J., Skillman, E.~D., et al.\ 2008, \mnras, 388, 1127. doi:10.1111/j.1365-2966.2008.13347.x
\bibitem[Herter et al.(2013)]{2013PASP..125.1393H} Herter, T.~L., Vacca, W.~D., Adams, J.~D., et al.\ 2013, \pasp, 125, 1393
\bibitem[Hofmann et al.(1995)]{1995A&A...300..403H} Hofmann, K.-H., Seggewiss, W., \& Weigelt, G.\ 1995, \aap, 300, 403
\bibitem[Hosokawa et al.(2010)]{2010ApJ...721..478H} Hosokawa, T., Yorke, H.~W., \& Omukai, K.\ 2010, \apj, 721, 478
\bibitem[Krumholz \& Tan(2007)]{2007ApJ...654..304K} Krumholz, M.~R. \& Tan, J.~C.\ 2007, \apj, 654, 304. doi:10.1086/509101
\bibitem[Lacy et al.(1982)]{1982ApJ...255..510L} Lacy, J.~H., Beck, S.~C., \& Geballe, T.~R.\ 1982, \apj, 255, 510. doi:10.1086/159851
\bibitem[Lim et al.(2016)]{2016ApJ...829L..19L} Lim, W., Tan, J.~C., Kainulainen, J., et al.\ 2016, \apjl, 829, L19. doi:10.3847/2041-8205/829/1/L19
\bibitem[Lim \& De Buizer(2019)]{2019ApJ...873...51L} Lim, W. \& De Buizer, J. M.\ 2019, \apj, 873, 51 [Paper I]
\bibitem[Lim et al.(2020)]{2020ApJ...888...98L} Lim, W., De Buizer, J.~M., \& Radomski, J.~T.\ 2020, \apj, 888, 98 [Paper II]
\bibitem[Ma{\'\i}z Apell{\'a}niz et al.(2022)]{2022A&A...657A.131M} Ma{\'\i}z Apell{\'a}niz, J., Barb{\'a}, R.~H., Fern{\'a}ndez Aranda, R., et al.\ 2022, \aap, 657, A131. doi:10.1051/0004-6361/202142364
\bibitem[Melena et al.(2008)]{2008AJ....135..878M} Melena, N.~W., Massey, P., Morrell, N.~I., et al.\ 2008, \aj, 135, 878. doi:10.1088/0004-6256/135/3/878
\bibitem[Melnick et al.(1989)]{1989A&A...213...89M} Melnick, J., Tapia, M., \& Terlevich, R.\ 1989, \aap, 213, 89
\bibitem[Moffat(1983)]{1983A&A...124..273M} Moffat, A.~F.~J.\ 1983, \aap, 124, 273
\bibitem[Moffat et al.(1994)]{1994ApJ...436..183M} Moffat, A.~F.~J., Drissen, L., \& Shara, M.~M.\ 1994, \apj, 436, 183. doi:10.1086/174891
\bibitem[M{\"u}cke et al.(2002)]{2002ApJ...571..366M} M{\"u}cke, A., Koribalski, B.~S., Moffat, A.~F.~J., et al.\ 2002, \apj, 571, 366. doi:10.1086/339843
\bibitem[N{\"u}rnberger et al.(2002)]{2002A&A...394..253N} N{\"u}rnberger, D.~E.~A., Bronfman, L., Yorke, H.~W., et al.\ 2002, \aap, 394, 253. doi:10.1051/0004-6361:20021022
\bibitem[N{\"u}rnberger \& Stanke(2003)]{2003AA...400..223N} N{\"u}rnberger, D.~E.~A. \& Stanke, T.\ 2003, \aap, 400, 223. doi:10.1051/0004-6361:20021894
\bibitem[N{\"u}rnberger(2003)]{2003A&A...404..255N} N{\"u}rnberger, D.~E.~A.\ 2003, \aap, 404, 255. doi:10.1051/0004-6361:20030453
\bibitem[O'dell et al.(1993)]{1993ApJ...410..696O} O'dell, C.~R., Wen, Z., \& Hu, X.\ 1993, \apj, 410, 696. doi:10.1086/172786
\bibitem[Retallack \& Goss(1980)]{1980MNRAS.193..261R} Retallack, D.~S. \& Goss, W.~M.\ 1980, \mnras, 193, 261. doi:10.1093/mnras/193.2.261
\bibitem[Robitaille et al.(2007)]{2007ApJS..169..328R} Robitaille, T.~P., Whitney, B.~A., Indebetouw, R., et al.\ 2007, \apjs, 169, 328. doi:10.1086/512039
\bibitem[Rochau et al.(2010)]{2010ApJ...716L..90R} Rochau, B., Brandner, W., Stolte, A., et al.\ 2010, \apjl, 716, L90. doi:10.1088/2041-8205/716/1/L90
\bibitem[R{\"o}llig et al.(2011)]{2011A&A...525A...8R} R{\"o}llig, M., Kramer, C., Rajbahak, C., et al.\ 2011, \aap, 525, A8. doi:10.1051/0004-6361/201014765
\bibitem[Sher(1965)]{1965MNRAS.129..237S} Sher, D.\ 1965, \mnras, 129, 237. doi:10.1093/mnras/129.3.237
\bibitem[Stolte et al.(2006)]{2006AJ....132..253S} Stolte, A., Brandner, W., Brandl, B., et al.\ 2006, \aj, 132, 253. doi:10.1086/504589
\bibitem[Sung \& Bessell(2004)]{2004AJ....127.1014S} Sung, H. \& Bessell, M.~S.\ 2004, \aj, 127, 1014. doi:10.1086/381297
\bibitem[Wang \& Chen(2010)]{2010SCPMA..53S.271W} Wang, J. \& Chen, Y.\ 2010, Science China Physics, Mechanics, and Astronomy, 53, 271. doi:10.1007/s11433-010-0052-y
\bibitem[We{\ss}mayer et al.(2023)]{2023arXiv230806164W} We{\ss}mayer, D., Przybilla, N., Ebenbichler, A., et al.\ 2023, \aap, 677, A175. doi:10.1051/0004-6361/202347253
\bibitem[Young et al.(2012)]{2012ApJ...749L..17Y} Young, E.~T., Becklin, E.~E., Marcum, P.~M., et al.\ 2012, \apjl, 749, L17
\bibitem[Zhang \& Tan(2011)]{2011ApJ...733...55Z} Zhang, Y., \& Tan, J.~C.\ 2011, \apj, 733, 55
\bibitem[Zinnecker \& Yorke(2007)]{2007ARA&A..45..481Z} Zinnecker, H. \& Yorke, H.~W.\ 2007, \araa, 45, 481. doi:10.1146/annurev.astro.44.051905.092549


\end{thebibliography}
\end{document}